\documentclass[aps,twocolumn,longbibliography,pra,superscriptaddress,amssymb,amsmath,floatfix]{revtex4-1}
\usepackage{amsmath}
\usepackage{graphicx}
\usepackage{epstopdf}
\usepackage{comment}
\usepackage{changepage}
\usepackage{bm}
\usepackage{wrapfig}
\usepackage{times}
\usepackage[colorlinks=true,linkcolor=blue,urlcolor=blue,citecolor=blue]{hyperref}
\usepackage{float}
\usepackage{titlesec}
\newcommand{\bea}{\begin{eqnarray}}
\usepackage{csquotes}
\usepackage{subfig}
\usepackage{xcolor}
\usepackage{tabularx}
\newcommand{\eea}{\end{eqnarray}}
\newcommand{\ninej}[9]{
\ensuremath{
\left\{\!\!
\begin{array}{ccc}
#1 & #2 & #3 \\
#4 & #5 & #6 \\
#7 & #8 & #9 \\
\end{array}
\!\!\right\}
}}

\begin{document}

\title{Multichannel Quantum Defect Theory with Numerical Reference Functions: Applications to Cold Atomic Collisions}

\author{Dibyendu Sardar}
\email{chem.dibyandu.sardar@gmail.com}
\affiliation{School of Physical Sciences, Indian Association for the Cultivation of Science, Kolkata 700032, India}
\author{Arpita Rakshit}
\affiliation{
$^{2}$ Engineering Institute for Junior Executives, Howrah, Department of Technical Education, Training and Skill Development, Govt  of WB, India}

\author{Somnath Naskar}
\affiliation{School of Physical Sciences, Indian Association for the Cultivation of Science, Kolkata 700032, India}
\affiliation{Department of Physics, Jogesh Chandra Chaudhuri College, Kolkata 700033, India; sommm147@gmail.com}
\author{Bimalendu Deb}
\affiliation{School of Physical Sciences, Indian Association for the Cultivation of Science, Kolkata 700032, India}

\date{\today}
\begin{abstract}
We develop a method for calculating multichannel wavefunctions in the spirit of quantum defect theory, based on numerically calculated reference functions. We benchmark the method by calculating cold collisional properties of $^{85}$Rb and $^6$Li in the presence of external magnetic fields tuned across specific $s$-wave Feshbach resonances and thereby reproducing known results. We then apply the method to calculate experimentally observed $d$-wave Feshbach resonance {[\textit{Phys. Rev. Lett.} {\bf 119}, 203402 (2017)]} 
 in $^{87}$Rb-$^{85}$Rb cold collisions. Our numerical results for this $d$-wave resonance show good agreement with the experimental observations. The method is applicable to arbitrary interaction potentials and to any energy range near the scattering threshold. The implementation of our method to any multichannel two-body scattering problem is straightforward.

\end{abstract}

\maketitle
\section{Introduction}
\label{sec:intro}
In recent times, cold atomic collisions have emerged as a key area of research, opening prospects for hither-to-unexplored regimes of cold chemistry \cite{balakrishnan:jcp:2016,coldchemistry}. One of the important methods for controlling the cold collisional properties of atoms is the magnetic Feshbach resonance (MFR) \cite{mfr,kohler2006}. MFR has remained an indispensable tool for tuning the $s$-wave scattering length \cite{ketterle:nature:1998}, and has been widely utilized in demonstrating a number of few- and many-body quantum effects using cold atomic gases \cite{bloch:rmp:2008,mfr}. Apart from MFR, an optical method, known as optical Feshbach resonance (OFR) \cite{fedichev:prl:1996,lett:prl:2002,deb:prl:2009,takahashi2008} that makes use of photoassociative coupling, is currently being explored as an alternative tool for controlling interatomic interactions.  Photoassociation (PA) \cite{jones:rmp:2006,wiener:rmp:1999} is a photochemical process by which a pair of colliding cold atoms becomes bound into a molecule in an electronically excited state by a single photon absorption. PA in the presence of an MFR can lead to Fano resonances \cite{deb1,li2}, which is a manifestation of quantum interference in spectroscopy or quantum collisions. At a fundamental level, all these resonances, namely, MFR, OFR, and Fano resonances, can be treated as a multichannel quantum scattering problem. Therefore, it is important to develop an accurate but preferably simple and robust numerical method for solving a generic multichannel scattering problem.

In 
 general, multichannel scattering problems are formulated through coupled Schr\"odinger equations, which may be treated using either a time-dependent or a time-independent formalism. The time-dependent approach is particularly useful for studying quantum reactive scattering processes \cite{zhao:pccp:2024,zhao:jctc:2024} at low collision energies, where dynamical information about the reaction pathways is of primary interest. However, in the ultracold regime---which is the main focus of the present study—the time-independent formalism is more commonly employed, as it provides direct access to scattering observables such as phase shifts, cross sections, and resonance properties with comparatively lower computational cost.

The most accurate {time-independent} numerical method for solving multichannel quantum scattering problems is the close-coupling (CC) method \cite{burke:rmp:1962,tamura,hutson94}. In this method, one needs to propagate wave functions in a matrix form outward starting from a short-range separation, and then match the functions with asymptotic boundary conditions.  For an $N$-channel CC problem, each step of propagation requires an $O(N^3)$ matrix operation, and so the usual CC algorithm takes a time proportional to $N^3$. Therefore, CC  calculations are computationally demanding. However, the properties of atomic and molecular collisions can also be calculated by several other methods, including asymptotic bound-state model (ABM) \cite{tiecke,wille} and multichannel quantum defect theory (MQDT) \cite{seaton}. In the ABM method, one calculates only the bound states close to the thresholds of the channel potentials to describe Feshbach resonance and scattering lengths, bypassing the computation of explicit scattering states \cite{tiecke,wille}.  Another important method is MQDT, which is the object of \mbox{this study. }

Historically, quantum defect theory (QDT) was formulated by Seaton \cite{seaton} in the context of atomic Rydberg spectroscopy and collisions, based on a systematic separation of short- and long-range Coulomb interactions. The framework was subsequently generalized to include attractive and repulsive charge–dipole ($\pm 1/r^2$) interactions \cite{greene,greene1} and polarization potentials \cite{watanable,book}
. Since then, QDT and its multichannel extension (MQDT) have been successfully applied to a broad class of scattering and spectroscopic problems, including negative-ion photodetachment \cite{watanable}, predissociation of atom–diatom van der Waals complexes \cite{raoult,raoult1}, and hyperfine interactions in molecular ions \cite{osterwalder:jcp:2004}. The prototype atom–molecule collision system Mg+NH has also been treated within the MQDT framework \cite{croft}. More recently, MQDT has been extended to describe low-energy atom–molecule chemical reactions with quantum-state resolution \cite{jisha-pra} and ro-vibrational transitions in ultracold molecule–molecule collisions \cite{jisha2014}. Alongside this, MQDT has been used extensively in the study of ultracold atomic collisions \cite{mies,mies1,julienne,burke,mies2,mies3,tiesinga,pires,hanna}. MQDT has also proven effective for ion–atom cold collisions \cite{idziaszek1,idziaszek2}, where the long-range interaction behaves as $-C_4/r^4$, in contrast to the $-C_6/r^6$ van der Waals potential relevant for neutral atom–atom systems. Analytical solutions of these long-range potentials \cite{gao-analytical} have been used \cite{gao1998,gao2010} to develop analytic MQDT formulations, yielding significant insight into ultracold atom–atom \cite{gao2006,gao2009,ruzic,you2017} and atom–ion \cite{gao2010,gao,li} collision physics.

MQDT starts by propagating outward the multichannel wave function or its log-derivative in a matrix form from short separations. But, instead of propagating into the asymptotic region as in CC, the propagation is stopped at an intermediate and classically allowed region, known as the matching point $r_m$. Then, this numerically obtained solution at $r_m$ is matched with the analytical solutions of the long-range form of the potential \mbox{matrix \cite{gao-analytical}} {or by numerically calculated reference functions} \cite{jisha-pra,jisha2014,ruzic}. Implicit assumptions or conditions in this method are that (i) for separations $r > r_m$,  the off-diagonal elements of the potential matrix (which are basically inter-channel couplings) become negligible, rendering the potential matrix essentially into a diagonal form, and  (ii) the diagonal elements of the potential matrix for $r > r_m$ should be expressible,  at least in leading order,  in analytical forms that should also admit analytical solutions. The latter condition is the most stringent one, restricting the application of MQDT methods to certain specific classes of long-range potentials, such as Coulomb ($1/r$), dipolar ($1/r^2$),   van der Waals ( $1/r^6$),  charge-neutral ($1/r^4$), etc. 

In this work, we develop a fully numerical MQDT framework that avoids these restrictions and is applicable to arbitrary long-range interaction potentials. Our approach proceeds in two main steps. First, we compute a pair of numerically exact reference functions for each channel by solving the single-channel Schrödinger equation using the Numerov–Cooley method \cite{johnson1977}, imposing physically appropriate asymptotic boundary conditions for open and closed channels. A second, linearly independent reference function is generated by solving the associated Wronskian equation, ensuring strict linear independence over the entire radial domain. These reference functions are then used to construct quantum defect functions, which are matched to the outward-propagated multichannel solution at a suitably chosen matching radius in the classically allowed region. This matching yields a short-range ${\bf R}$ 
 matrix \cite{seaton} that fully characterizes the interaction region. In the second step, asymptotic analysis is performed to obtain physically acceptable solutions and extract scattering phase shifts. {In this context, it is worth pointing out that semi-analytical or numerical reference functions  were previously employed by other authors \cite{jisha-pra,jisha2014,ruzic} in multichannel  scattering problems. One important feature of  our approach  is to evaluate a pair of accurate  reference functions by numerically solving  Wronskian equations for all channels including the closed ones.  Our method does not require calculating explicitly the exponentially growing solutions of the closed channels in the entire range of classically forbidden region. In fact, we calculate the closed channel solutions for a limited range close to the classically allowed region. Our method is designed in such a way that these solutions are accurate and the implementation of the method is straightforward} \cite{debbook2026}.

We apply this method to ultracold atomic collisions and the calculation of magnetic Feshbach resonances. As benchmarks, we reproduce well-known $s$-wave resonances in $^{85}$Rb+$^{85}$Rb \cite{nygard,burnett} and 
 $^{6}$Li+$^{6}$Li \cite{mfr} systems. We then extend the analysis to higher partial-wave resonances, focusing on an experimentally observed broad $d$-wave Feshbach resonance in the $^{85}$Rb+$^{87}$Rb system \cite{you2017}. Using a full six-channel numerical MQDT calculation and high-precision interaction potentials \cite{strauss2010}, we locate the broad resonance near 424.1~G at a collision energy of {37~$\mu$K,} 
 in close agreement with experiment. At lower energies, inclusion of second-order spin–spin interactions leads to a characteristic triplet splitting associated with the $\lvert m_\ell \rvert = 0,1,2$ sublevels of the $d$-wave rotational angular momentum, in qualitative agreement with experimental observations.

In comparison with other MQDT approaches, the present method offers several notable advantages. First, it is applicable to arbitrary forms of long-range interaction potentials, in contrast to MQDT formulations that rely on explicit analytical solutions valid only for selected classes of long-range potentials. Second, the method does not require WKB-type boundary conditions, as commonly employed in semi-analytical MQDT treatments, for the construction of reference-function pairs. Third, the numerically generated reference functions are guaranteed to remain strictly linearly independent over the entire spatial domain of interest. Fourth, the availability of such linearly independent reference functions allows for the straightforward construction of the real-space Green’s function. As a result, the effects of any residual potential matrix can be incorporated as final-state interactions within a perturbative framework, thereby enabling more accurate calculations of multichannel scattering wave functions. 

The remainder of the paper is organized in the following way. In Section \ref{sec:level1}, we describe our numerical method for the calculation of reference functions.  In Section \ref{sec:level5}, we verify our method by reproducing scattering properties for a two-channel model potential of the $^{85}$Rb system anda  five-channel $^6$Li system.  Section \ref{sec:level8} describes the application of our method to calculate higher partial-wave MFR.  Finally, we make conclusions in Section \ref{sec:level20}.



\section{Theoretical Approach: 
 Numerical Reference Functions}\label{sec:level1}
In this section, we describe the MQDT prescription with a complete numerical approach based on numerical reference functions. A multichannel wave function for the $i$-th incident channel is expressed in the form  
\bea 
\mid\Phi_{i}\rangle =
\sum_{j} F_{j i} (r) \mid j \rangle,
\eea 
where $\mid j \rangle$ represents the channel babis state, and $F_{j i} (r)$ is the corresponding radial wave-function component. In the absence of any magnetic field, for a pair of ground-state atoms $a$ and $b$, a channel state $\mid j \rangle$ is defined as $\mid j \rangle \equiv \mid (f_a f_b), f, \ell, J \rangle$, where $f_{a(b)}$ is the hyperfine quantum number of the atom $a$($b$), $f =  f_a +  f_b$ and $\ell$ denotes the angular momentum (partial wave) of relative motion. Here, ${\vec J} = {\vec f} + \vec{\ell}$ is the total angular momentum. In the absence of any external magnetic field,  $f$, $\ell$, and $J$ are good quantum numbers. In the presence of an external magnetic field, the projection of the total angular momentum $M_J = M_F + m_\ell$ along the quantization axis remains a good quantum number. Here, \mbox{$M_F = m_{s_a} + m_{i_a} + m_{s_b} + m_{i_b}$} is the projection of the total atomic spin, where $m_{s_{a(b)}}$ and $m_{i_{a(b)}}$ denote the projections of the electronic spin $s_{a(b)}$ and nuclear spin $i_{a(b)}$ of atom $a(b)$, respectively. The quantity $m_\ell$ is the projection of the relative orbital angular momentum $\vec\ell$. Here, we have assumed that the rotational motion of the internuclear axis is uncoupled or weakly coupled with the internal spin motion.  
In that case, a channel is defined by diagonalizing the Hamiltonian of two non-interacting atoms including the atomic Zeeman shifts,  resulting in a channel state which is a superposition of the product angular momentum states $\mid (s_a i_a, m_{s_a} m_{i_a}); (s_b i_b, m_{s_b} m_{i_b}) \rangle \otimes \mid \ell m_{\ell}  \rangle$. 

The full multichannel wave function can be conveniently expressed in a matrix form $\mathbf{\Psi}(r)$, whose elements are $F_{j i}(r)$. The coupled radial Schr\"{o}dinger equations are then given by \cite{croft} 
\bea
\left[-\frac{\hbar^2}{2\mu}\frac{d^2}{dr^2}-E\right] F_j(r)+\sum_{i}W_{ij}(r)F_i(r)=0,
\label{coupled-eq}
\eea
where $E$ is the collision energy, $\mu$ is the reduced mass, and the coupling matrix elements are \cite{croft}

\begin{equation}
\begin{split}
W_{ji}(r)
&=
\int \chi_j^{*}(\tau)
\Biggl[
\hat H_{\mathrm{int}}(\tau)
+V(r,\tau) \\
&\qquad
+\frac{\hbar^2\ell_i(\ell_i+1)}
{2\mu r^2}
\Biggr]
\chi_i(\tau)\,d\tau .
\end{split}
\end{equation}

Here, 
   $\tau$ represents an internal degree-of-freedom of the system. At large interatomic separation, the potential matrix becomes asymptotically diagonal \cite{croft},
\bea
W_{ji}({r\rightarrow\infty})\sim\left[E_{i}^{\infty}+\frac{\hbar^2\ell_i(\ell_i+1)}{2\mu r^2} \right]\delta_{ij}+O(r^{-n}),
\label{asmp}
\eea
where $\ell_i$ denotes the partial wave, and $E_i^{\infty}$ is the threshold energy of the $i$-th channel, $n$ is the power of the leading term in the potential expansion. In matrix representation \cite{croft}, the Equation (\ref{coupled-eq}) becomes
\bea
-\frac{\hbar^2}{2\mu}\frac{d^2 \mathbf{\Psi}}{dr^2}=\left[{\bf W}(r)-E{\bf I}\right] \mathbf{\Psi}(r),
\label{ce}
\eea   
where $\mathbf {I}$ is the identity matrix. 

For the calculation of scattering states and bound states, the wave function should be regular at the origin. As $r\rightarrow0$, $V(r)>>0$, and the short-range boundary condition is
\bea
F_i(r)\rightarrow 0\hspace{0.1cm}; \hspace{1.2cm} r\rightarrow0
\eea

For an $N$-channel problem, the above coupled equations yield $N$ solution vectors that should satisfy the boundary condition at $r\rightarrow 0$ and form an $(N \times N )$ radial wave functions matrix $\mathbf{\Psi}(r)$ \cite{croft}.

\subsection{\label{sec:level2}Inward Propagation: Solving Wronskian Equation}

Let there be $N_o$ number of open channels, which are enumerated starting from 1 to $N_o$, and 
$N_c$ number of closed channels from $N_o + 1$ to $N$  with total channels being  $N = N_o + N_c$. Initially, we calculate one solution $\phi_i(r)$ of each channel $i$  
by inward propagation of a standard single-channel Numerov-Cooley code starting from the asymptotic limit up to the matching point $r_m$.  Since the inter-channel mixing is negligible in this domain, independent single-channel propagation can be pursued.  For a closed channel $i$, the asymptotic boundary condition is set as   
\bea
\phi_i(r\rightarrow\infty)\sim \exp(-\kappa_ir);  \hspace{0.2in} i=N_o + 1, \cdots  N 
\eea
 where $\kappa_i = \sqrt{W_{ii}(r\rightarrow \infty) - k^2}$ with $k^2 = 2 \mu E /\hbar^2$.
The point $r_m$ is chosen in a classically allowed region where the wavefunction $\phi_i(r)$ crosses the first anti-node from the outer side. For the open channels from $i= 1 $ to $i=N_o$, we consider a sinusoidal asymptotic boundary condition $\phi_i(r) \sim \sin (k_i r - \ell \pi/2)$ where $k_i = \sqrt{k^2 - W_{ii}(r\rightarrow \infty)}$. 
Once the function $\phi_i(r)$ for each channel $i$ is found, we calculate another solution $\psi(r)$ by solving the Wronskian equation $\phi'_i(r)\psi_i(r)-\psi'_i(r)\phi_i(r)=C$, where $C = k_i$ if the channel is open or $C = - 2 \kappa_i$ if it is closed. 
 Although the Wronskian equation (which is a first-order inhomogeneous equation) admits an analytical solution \cite{arfken}, it is not of much use in practice as it can lead to numerical instability at or near the nodal points of $\phi_i(r)$. Instead, we solve this equation numerically to find the second solution $\psi_i(r)$.  The numerical procedure for solving the Wronskian equation is discussed in Appendix \ref{sec:level19}.

For a closed channel $i$, we make linear combinations of these two linearly independent functions to obtain two new linearly independent functions that asymptotically go as  sine  and cosine hyperbolic functions.   
Let us denote this pair of functions as $s_{c_i}$ and $c_{c_i}$ 
\bea
s_{c_i}(r)= n_i (\phi_i(r)-\psi_i(r)),
\eea
\vspace{-24pt}
\bea
c_{c_i}(r)=n_i (\phi_i(r)+\psi_i(r)),
\eea
where $n_i = \sqrt{\kappa_i/\pi |E_i|}$ is the normalization constant (for energy normalization) with $E_i = - \hbar^2 \kappa_i^2/2\mu$ being the asymptotic closed-channel energy.  
For an open channel $i$, the corresponding pair functions are obtained by normalizing the functions $\phi_i(r)$ and $\psi_i(r)$ with the normalization constant  $n_i = \sqrt{k_i/\pi E_i}$ with $E_i = \hbar^2 k_i^2/2\mu$.
Thus, we obtain a desirable linearly independent energy-norma zed  base pair or reference functions for both open and closed channels for building up our MQDT.

\subsection{\label{sec:level3}Outward Propagation}
Next, we calculate wave functions in matrix form by performing outward
propagation from $r \sim 0$ to $r_m$, considering the short-range boundary condition. For the $N_c$ number of closed channels, the radial functions will, in general, be exponentially rising in the large limit of $r$. But, the physical solutions should be bounded everywhere. Accordingly, we define a matrix $\mathbf{G}(r)$ whose closed-channel components vanish asymptotically \cite{seaton},
\bea
{G}_{ij}({r\rightarrow\infty})\sim 0, \hspace{0.4in}  \text{where} \hspace{0.2in} i=N_o+1 \hspace{0.1in}{\text to} \hspace{0.1in}N_c.  
\eea

For 
 open channels, the asymptotic behavior of $\mathbf{G}(r)$ defines the scattering reactance matrix $\mathbf{R}$, and is given by 
\bea
G_{ij}({r\rightarrow\infty})\sim s_{i}\delta(i,j)+c_{i}{R}_{ij},  \hspace{0.05in} \text{where} \hspace{0.05in} i,j=1,... N_o.
\eea

In 
 matrix notation,
\bea
{\bf G}_{\mathrm{oo}}(r)\sim {\bf s} + {\bf c}{\bf R}, \hspace{0.4in} {\bf G}_{\mathrm{co}}(r)\sim 0, 
\label{eq13}
\eea
where $\mathbf s$ and $\mathbf c$ are diagonal matrices of open-channel reference functions, ${\bf G}_{\mathrm{oo}}$ and ${\bf G}_{\mathrm{co}}$ matrixes denote open--open and closed--open counterparts \cite{seaton}.

Finally, the outward propagated solutions are matched at $r_m$ to the inward reference functions via \cite{seaton}  
\bea
{\bf F}(r)={\bf s} + {\bf c} {\bf R},  \hspace{0.3in} \text {for}\hspace{0.15in} {r \geq} {r_\text{m}}.
\label{fr}
\eea

 
\subsection{\label{sec:level4}Elimination of Exponentially Growing Solutions of Closed Channels}
The physical solution matrix ${\bf G}(r)$ is related to ${\bf F}(r)$ by \cite{seaton}  
\bea
{\bf G}(r)= {\bf F}(r){\bf L},
 \label{gg}
\eea
where $\bf L$ is a column matrix, having $N_o$ columns and $N$ rows. Partitioning the matrices into open and closed subspaces,

\begin{equation}
\bf{F}=
\begin{pmatrix}
\bf{F}_{\mathrm{oo}} & \bf{F}_{\mathrm{oc}} \\
\bf{F}_{\mathrm{co}} & \bf{F}_{\mathrm{cc}}
\end{pmatrix},
\qquad
\bf{L}=
\begin{pmatrix}
\bf{L}_{\mathrm{oo}} \\
\bf{L}_{\mathrm{co}}
\end{pmatrix},
\end{equation}
we obtain
\begin{equation}
\bf{G}_{\mathrm{oo}}=
\bf{F}_{\mathrm{oo}}\bf{L}_{\mathrm{oo}}
+\bf{F}_{\mathrm{oc}}\bf{L}_{\mathrm{co}},
\end{equation}
\begin{equation}
\bf{G}_{\mathrm{co}}=
\bf{F}_{\mathrm{co}}\bf{L}_{\mathrm{oo}}
+\bf{F}_{\mathrm{cc}}\bf{L}_{\mathrm{co}}.
\end{equation}
where the sub-matrices ${\bf F}_{oo}$, ${\bf F}_{oc}$, ${\bf F}_{co}$ and ${\bf F}_{cc}$ are of dimensions $N_o\times N_o$, $N_o\times N_c$, $N_c\times N_o$ and $N_c\times N_c$, respectively. Here, dimensions of $\bf L_{oo}$ and $\bf L_{co}$ matrices are $N_o\times N_o$ and $N_c\times N_o$, respectively.
Comparison with the asymptotic form yields
\begin{equation}
\bf{L}_{\mathrm{oo}}=\bf{I},
\qquad
\bf{R}=
\bf{R}_{\mathrm{oo}}+
\bf{R}_{\mathrm{oc}}\bf{L}_{\mathrm{co}}.
\end{equation}

The condition $\bf{G}_{\mathrm{co}}={0}$ eliminates exponentially growing closed-channel solutions and leads to
\begin{equation}
\bf{L}_{\mathrm{co}}=
\big(\bf{I}-\bf{R}_{\mathrm{cc}}\big)^{-1}
\bf{R}_{\mathrm{co}}.
\end{equation}

Substituting this expression yields the final reactance matrix $\bf{R}$, which is equivalent to the physical scattering ${\bf K}$-matrix. The scattering ${\bf K}$-matrix is related to ${\bf S}$-matrix 
\begin{equation}
\bf{S}=
(\bf{I}-i\bf{K})^{-1}
(\bf{I}+i\bf{K}),
\end{equation}
and the transition ${\bf T}$-matrix is defined as
\begin{equation}
\bf{T}=\bf{I}-\bf{S}.
\end{equation}

 The open channel component of ${\bf T}$-matrix is related to the scattering phase shift \mbox{$(\delta)$ \cite{nygard}} by 
\begin{equation}
 \mid T\mid^2 = 4\sin^2\delta
\end{equation}

\section{Verification}\label{sec:level5}

\begin{figure*}[t!]
    \centering
    \centering {{\includegraphics[width=0.49\textwidth]{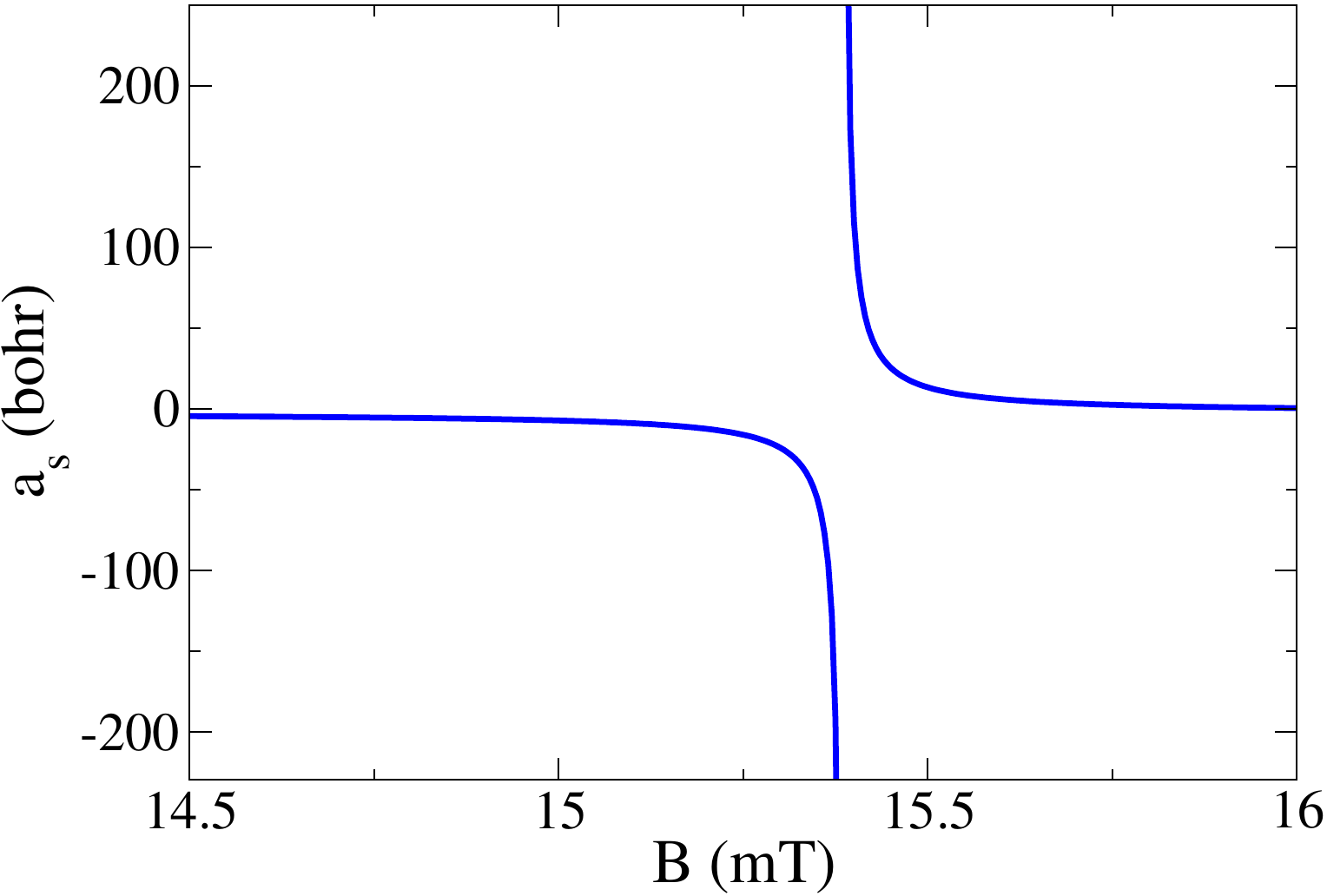} }}%
    \qquad
    \centering{{\includegraphics[width=0.445\textwidth]{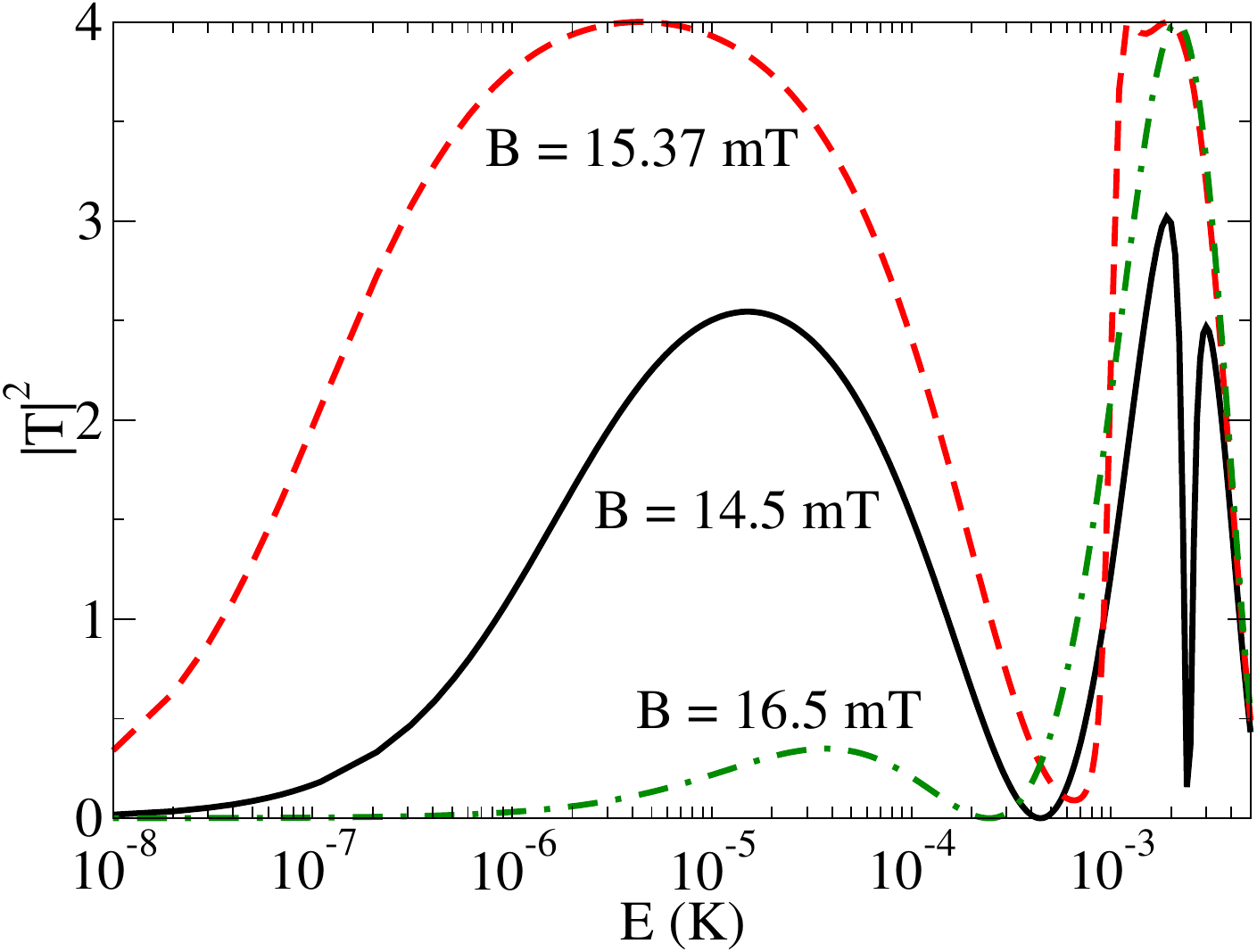} }}%
    \caption{Left hand panel shows the variation  of the $s$-wave scattering length $a_s$ of $^{85}$Rb as a function of magnetic field $B$ in mT. Right hand panel presents the variation of the absolute value of the square of T-matrix elements as a function of energy $E$ in Kelvin of $^{85}$Rb for three different values of magnetic field. }%
    \label{fig:res+tmat}%
\end{figure*}

To assess the validity and numerical accuracy of the method proposed above, we first apply it to reproduce several well-established $s$-wave Feshbach resonances. Having verified the performance of the method in this simplest and most thoroughly studied case, we then demonstrate its applicability to multichannel resonances involving higher partial waves by calculating a $d$-wave Feshbach resonance in the following section.

\subsection{\label{sec:level6}Feshbach Resonance for $^{85}\mathrm{Rb}+{}^{85}\mathrm{Rb}$}

We consider a previously studied two-channel model~\cite{nygard} describing a Feshbach resonance in collisions of $^{85}\mathrm{Rb}$ atoms. The Hamiltonian of the model is written as
\begin{equation}
\hat{H} =
\begin{pmatrix}
-\dfrac{\hbar^2}{2\mu}\dfrac{d^2}{dr^2} + V_{\mathrm{op}}(r) & W(r) \\
W(r) & -\dfrac{\hbar^2}{2\mu}\dfrac{d^2}{dr^2} + V_{\mathrm{cl}}(\vec{B},r)
\end{pmatrix},
\end{equation}
where $\mu$ is the reduced mass of the $^{85}\mathrm{Rb}+{}^{85}\mathrm{Rb}$ system. The diagonal elements describe the open and closed scattering channels, while $W(r)$ represents the inter-channel coupling.

The open-channel interaction potential $V_{\mathrm{op}}(r)$ is approximated by a Lennard--Jones potential of the form
\begin{equation}
V_{\mathrm{op}}(r)
=4\xi\left[\left(\dfrac{\sigma}{r}\right)^{12}
-\left(\dfrac{\sigma}{r}\right)^6\right],
\end{equation}
where $\sigma = 10.075\,a_0$ and $4\xi\sigma^6=C_6=4700\,a_0$. These parameters are taken from Ref.~\cite{nygard}. 

The closed-channel potential is assumed to have the same radial dependence as the open-channel potential, but with its asymptotic threshold shifted upward in energy by the Zeeman splitting,
\begin{equation}
V_{\mathrm{cl}}(r,\vec{B})
=V_{\mathrm{op}}(r)+E_{\mathrm{th}}+\Delta\mu B,
\end{equation}
where $\Delta\mu$ is the difference in magnetic moments between the separated atoms and the bare closed-channel bound state. Equivalently, we write $
V_{\mathrm{cl}}(r,\vec{B}) = V_{\mathrm{op}}(r)+E_{\mathrm{cl}}(\vec{B})$, where $E_{\mathrm{cl}}(\vec{B})$ follows the dependence of energy difference of the
corresponding Zeeman hyperfine levels with the magnetic field.

The coupling between the open and closed channels is modeled as
\begin{equation}
W(r)=\bar\beta\,\exp(-r/\alpha),
\end{equation}
with $\bar\beta=0.203$~a.u. and $\alpha=1\,a_0$. Together, the parameters $\sigma$, $\xi$, $\Delta\mu$, $\bar\beta$, and $\alpha$ define a five-parameter model that captures the essential physics of the $^{85}\mathrm{Rb}$ Feshbach resonance~\cite{nygard}.

Following the procedure described in Section~\ref{sec:level2} and Appendix~\ref{sec:level19}, we numerically construct the reference functions associated with the closed channel, namely the hyperbolic functions $s_{c_i}$ and $c_{c_i}$, as well as the sinusoidal reference functions $s_{o_i}$ and $c_{o_i}$ associated with the open channel. The inward propagation is performed from $r=100\,a_0$ for the closed channel and from $r=2000\,a_0$ for the open channel.
The coupled-channel equations are propagated outward in matrix form from $r=8\,a_0$ to a matching point $r_m$. The inward and outward solutions are matched using a two-point matching procedure in the spirit of quantum defect theory \cite{seaton},
${\bf\Phi}(r_m) =
\bigl({\bf s}(r_m)+{\bf c}(r_m)\bf{R}\bigr)\bf{A},
$
where $\bf{s}$ and $\bf{c}$ are diagonal matrices of reference functions, $\bf{R}$ is the short-range reactance matrix, and $\bf{A}$ is a normalization matrix. The matching point $r_m$ is chosen at the position where the inward-propagated reference function $\phi_i(r)$ crosses its first antinode when approaching from outer side. For the present two-channel model, this occurs near $r_m \approx 40\,a_0$, where the off-diagonal coupling is already negligible compared to the diagonal potentials. For a two-channel system, the reactance matrix reduces to a scalar $R$, and the scattering phase shift $\delta(k)$ is given by $\tan\delta(k)=R $. The $s$-wave scattering length then follows as: $a_s=-\lim_{k\rightarrow 0}\dfrac{\tan\delta(k)}{k}$.


Left hand panel of Figure~\ref{fig:res+tmat} shows the variation of the scattering length $a_s$ as a function of the magnetic field $B$. A resonance is observed near $B=15.4$~mT, in excellent agreement with the experimentally reported value of $15.5$~mT~\cite{burnett,nygard}. In the right hand panel of Figure~\ref{fig:res+tmat}, we plot the squared magnitude of the $T$-matrix as a function of collision energy for several values of $B$. Near the resonance field, $|T|^2$ approaches its maximum value of $4$, indicating that the scattering phase shift passes through $\pm\pi/2$, thereby confirming the occurrence of a Feshbach resonance.

Finally, we comment on the validity of neglecting the off-diagonal inter-channel coupling matrix elements $W_{ij}(r)$ ($i\neq j$) for $r>r_m$. To quantify the importance of residual coupling effects, we define the dimensionless ratio
\begin{equation}
\lambda
=
\dfrac{2W_{12}(r_m)}{W_{11}(r_m)+W_{22}(r_m)},
\end{equation}
which measures the relative strength of the off-diagonal coupling compared to the average diagonal potential at the matching point. When $\lambda\ll 1$, the neglect of inter-channel coupling for $r>r_m$ is well justified. For the present two-channel model, $\lambda$ is smaller than unity by several orders of magnitude, confirming the validity of our approximation.

\subsection{\label{sec:level8} $s$-Wave Feshbach Resonance of $^6$Li}
We reproduce and verify a known broad Feshbach resonance \cite{mfr} of fermionic $^6$Li atoms using our present method. We consider five asymptotic channels to reproduce the broad Feshbach resonance near 832 G. The Hamiltonian can be  written in the form 
\bea
H = T(r) + \sum {H^{int}} + V^c, 
\eea
where $T(r)$ is the kinetic energy term, $V^c$ is the interatomic potential  on electronic spin state $\vec{S}_1$ and $\vec{S}_2$ of the two atoms. The interaction may be written in the form of
\bea
V^c = V_0(r)P_0 + V_1(r)P_1, 
\eea
where $P_0 = 1/4 -{\vec S_1}\cdot{\vec S_2} $ and $P_1 = 3/4 + {\vec S_1}\cdot{\vec S_2} $ are the projection operators for two-electron singlet and triplet states, respectively; and $V_0(r)$ and $V_1(r)$ are the singlet and triplet potentials, respectively. This interaction is therefore diagonal in molecular or adiabatic basis $\mid I M_I; S M_S \rangle$, so that 
\begin{equation}
    \begin{split}
     \langle S' M'_S; I' M'_I \mid V^c \mid S M_S; I M_I \rangle &= \\\delta_{I,I'} \delta_{M_I M'_I} \delta_{S,S'} \delta_{M_S M'_S} V_S,   
    \end{split}
\end{equation}
where $\vec{S} = \vec{s}_1 + \vec{s}_2$ and $\vec{I} = \vec{i}_1 + \vec{i}_2$, $\vec{s}_1$ and $\vec{s}_2$ being the electronic spins and  $\vec{i}_1$ and $\vec{i}_2$ being nuclear spins of the two atoms. 
The interaction Hamiltonian $\sum{H^{int}}$ can be written  as
\bea
H^{int} = H_{hf} + H_{B}, 
\eea
where, $H_{hf}$ and $H_{B}$ represent hyperfine and Zeeman interactions, respectively. 

\vspace{-6pt}
\begin{table}[H]
\caption{Separated five atomic channels for the $s$-wave Feshbach resonance of $^6$Li. The projection of total angular momentum $M_F=0$. }
\begin{ruledtabular}
\begin{tabular}{c c c} 
 \text{Channels} & {\text{$(f_1, f_2)$}} & {\text{$(m_{f_1}, m_{f_2})$ }} \\ 
 \hline
 ab & $(\frac{1}{2}, \frac{1}{2})$ & $(+\frac{1}{2}, -\frac{1}{2})$  \\ 
 ad & $(\frac{1}{2}, \frac{3}{2})$ & $(+\frac{1}{2}, -\frac{1}{2})$  \\
 be & $(\frac{1}{2}, \frac{3}{2})$ & $(-\frac{1}{2}, +\frac{1}{2})$  \\
 cf & $(\frac{3}{2}, \frac{3}{2})$ & $(-\frac{3}{2}, +\frac{3}{2})$  \\ 
 de & $(\frac{3}{2}, \frac{3}{2})$ & $(+\frac{1}{2}, -\frac{1}{2})$  \\
\end{tabular}
\label{channel}
\end{ruledtabular}
\end{table}

\begin{figure*}[t]
    \centering
    \centering {{\includegraphics[width=0.475\textwidth]{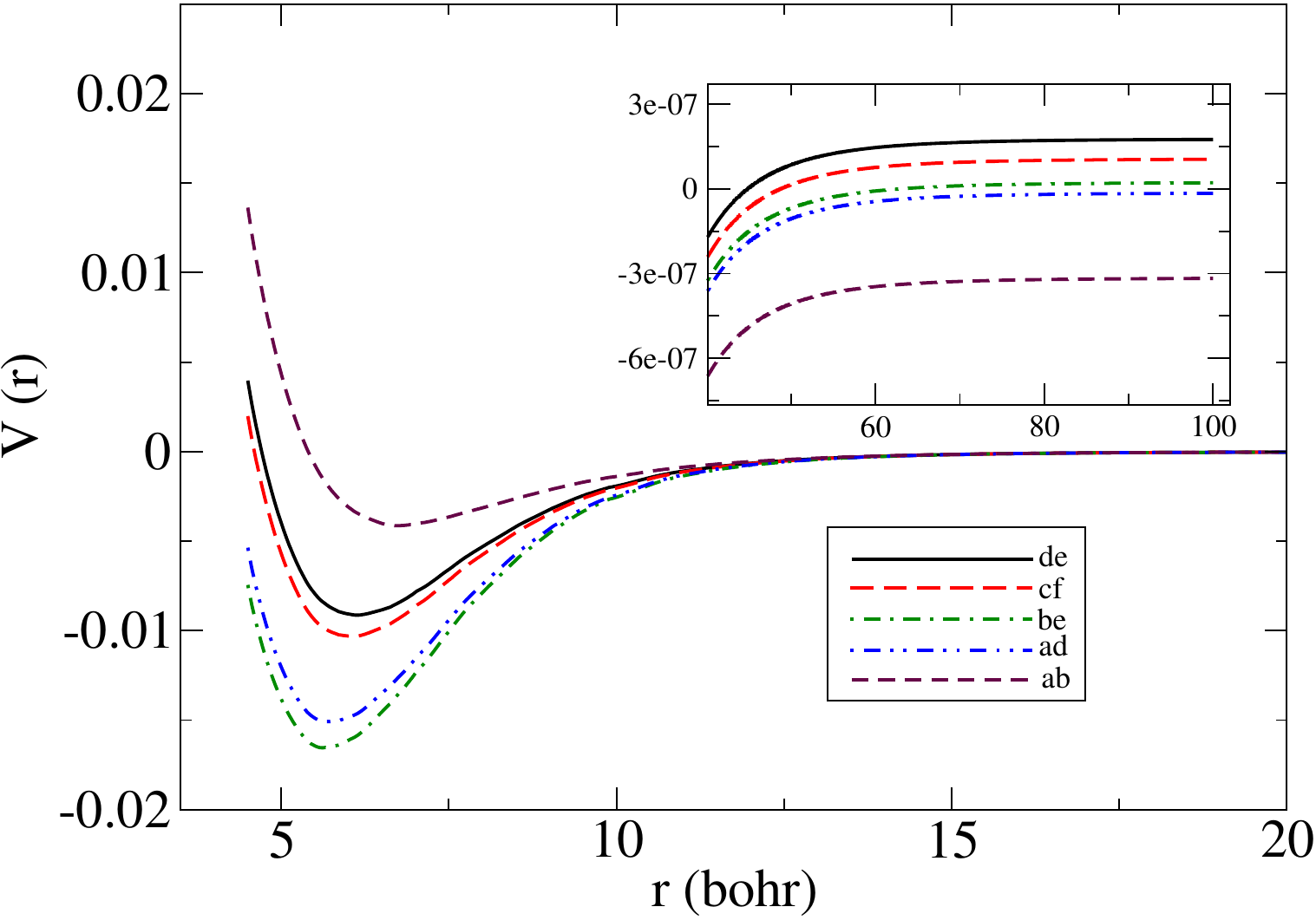} }}%
    \qquad
    \centering{{\includegraphics[width=0.46\textwidth]{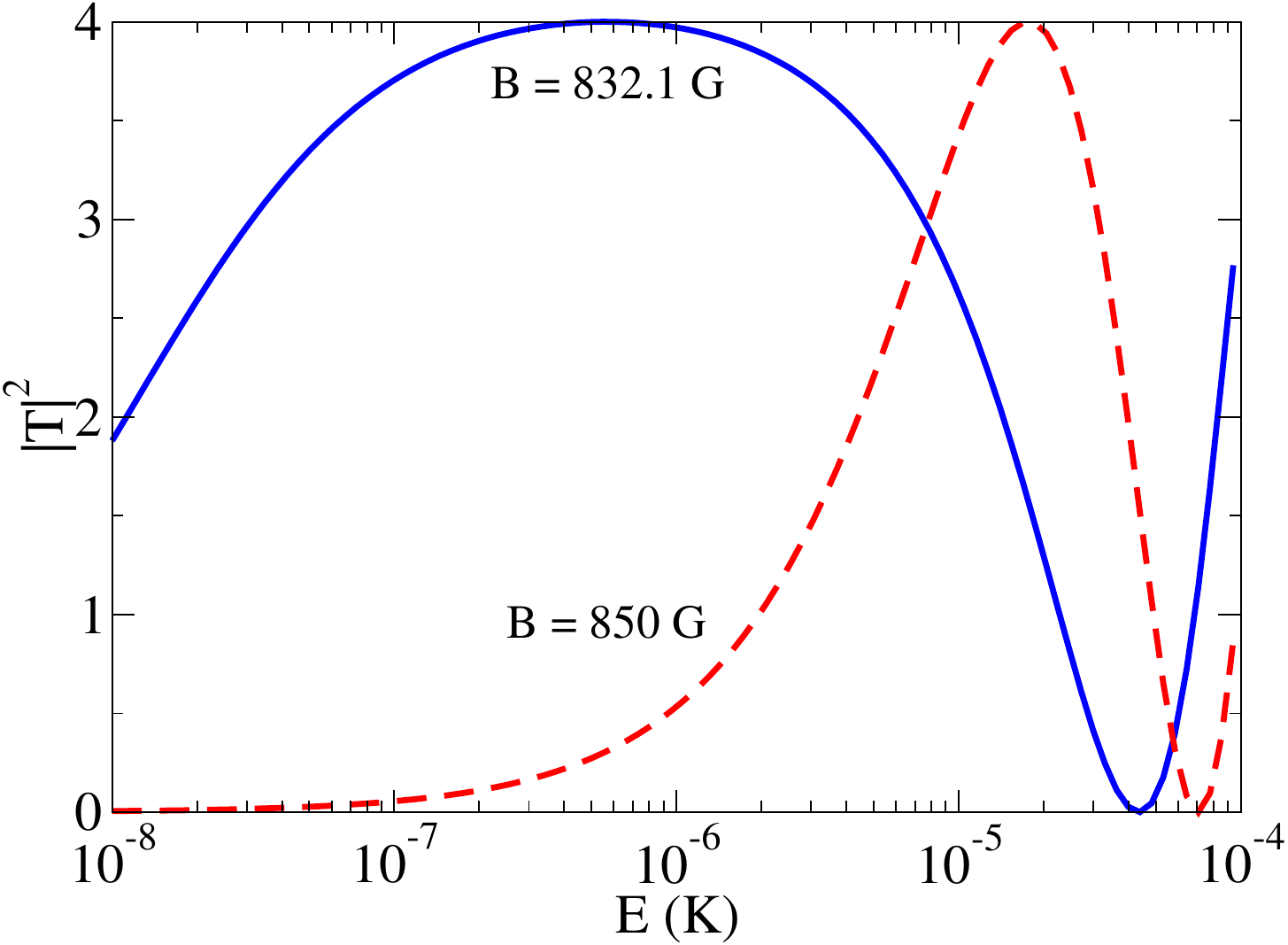} }}%
    \caption{Left hand panel shows the variations of five diagonal potentials of a $^6$Li system as a function of internuclear distance $r$ in the short-range regime 
for the magnetic field $B=832.1$ G. The corresponding asymptotes are shown in the inset of the figure. Right hand panel presents the variation of $|T|$$^2$ as a function of $E$ in Kelvin for two different values of $B=832.1$ G (blue, solid) and $B=850$ G (red, dashed) }%
    \label{fig:pot+tmat}%
\end{figure*}

Now, when the two atoms are well separated,  the non-interacting atoms can be treated individually in terms of an atomic basis $\mid f_j m_j \rangle$ for atom $j$, where $ \vec{f}_j = \vec{s}_j + \vec{i}_j$ and $m_j$ is the projection of the total spin for a single atom. The hyperfine interaction for a single atom can be written as 
\begin{equation}
    \begin{split}
     H_{hf} &=\frac{a_{hf}}{\hbar^2} \vec{s_j}\cdot\vec{i_j} 
     =\frac{a_{hf}}{2\hbar^2}(\vec{f_j}^2-\vec{s_j}^2-\vec{i_j}^2)\\
    &=\frac{a_{hf}}{2\hbar^2}[f_j(f_j+1)-s_j(s_j+1)-i_j(i_j+1)],
    \end{split}
\end{equation}
 where and $a_{hf}$ is the hyperfine constant.
 
Hence, for two colliding atoms at large separation, the suitable representation would be an uncoupled hyperfine basis $\mid f_1 m_1, f_2 m_2 \rangle$. So in the absence of a magnetic field, the interaction Hamiltonian can be written as  $\sum{H^{int}} = H_{hf} = H_1^{hf} + H_2^{hf}$. The atomic or diabatic or long-range basis can also be expressed in coupled hyperfine representation $\mid (f_1 f_2)F m_F \rangle$ and the hyperfine interaction is diagonal in this basis. Here, $ F = f_1 + f_2$ is the total hyperfine spin, and  $m_F$ is the projection of the total hyperfine spin. Now, we have to convert the central potential in the diabatic basis, $\mid (f_1 f_2)F m_F  \rangle$. 
\begin{equation}
    \begin{split}
     \langle (f_1 f_2) f m_f \mid V^c \mid (f'_1 f'_2) f' m'_f \rangle &= \\\sum_{S,I,M_S,M_I} V_S\langle (f_1 f_2) f m_f; \ell m_\ell \mid S M_S; I M_I; \ell' m_\ell'  \rangle\\ \langle S M_S; I M_I; \ell' m'_\ell \mid (f_1 f_2) f m_f; \ell m_\ell \rangle.    
    \end{split}
\end{equation}

The 
 transformation of the diabatic basis (coupled hyperfine representations) to the adiabatic basis (short range representations) is as follows \cite{burke1999theoretical}
 \vspace{-6pt}
\begin{equation}
    \begin{split}
    \langle S M_S; I M_I; \ell' m'_\ell \mid (f_1 f_2) f m_f; \ell m_\ell \rangle = \\ \delta_{\ell \ell'}  \delta (m_\ell m'_\ell) \langle S M_S; I M_I \mid f  m_f \rangle \\  \sqrt{(2 f_1 + 1)(2 f_2 + 1)(2 S + 1)(2 I + 1)}\\
\ninej{s_1}{i_1}{f_1}{s_2}{i_2}{f_2}{S}{I}{f}
\left( \frac{1+(1-\delta_{f_1 f_2})(-1)^{S+I+l}}{\sqrt{2-\delta_{f_1 f_2}}}\right).    
    \end{split}
\end{equation}

Here, 
 $\langle S M_S; I M_I \mid f  m_f \rangle$ is the Clebsch--Gordan coefficient \cite{burke1999theoretical}  and the quantity in curly brakets is known as  $9j$-symbol. Here $m_1 + m_2 = m_f = m'_1 + m'_2 = M_S + M_I$ 

If the magnetic field is sufficiently weak, then as a first approximation, one can use these channel states. However, when the magnetic field is strong enough, the asymptotic Hamiltonian is no longer diagonal due to the presence of Zeeman terms.  A new basis denoted by $ {\mid  {\tilde f \tilde m_f} \rangle}$, which is suitable for scattering in the presence of a magnetic field, is obtained by diagonalizing the asymptotic form of the Hamiltonian. But, in this new basis, the central potential $V^c$ can not be diagonalized, and the resulting off-diagonal terms will provide the coupling which may eventually lead to multichannel resonances. Let \mbox{$\mid a\rangle=\mid S M_S; I M_I\rangle$} denote an adiabatic basis that diagonalizes $V^c$; 
$\mid b\rangle=\mid f_1 m_1; f_2 m_2\rangle$ is an asymptotic basis that diagonalizes $H_{hf}$ and $\mid{\tilde b}\rangle= {\mid  {\tilde f \tilde m_f} \rangle}$ diagonalizes $H_{hf}+H_B$, respectively. So, we need to express the whole problem in $\mid{\tilde b}\rangle$ basis, which is a physically 
relevant basis for our 
purpose at $r\rightarrow\infty$. We consider the following steps in order to obtain the diagonal and off-diagonal potentials in the physically relevant basis. In the first step, we express $(H_{hf}+H_B)$ in $\mid b\rangle$ basis, which can be done analytically and leads to a non-diagonal matrix. Then, this matrix is numerically diagonalized to obtain eigenvalues which define the threshold energy of the channel and the eigenvectors for transformations from $\mid b\rangle$ to $\mid{\tilde b}\rangle$ basis.
\bea
\mid{\tilde b_j}\rangle=\sum_i \mid b_i\rangle \langle b_i \mid{\tilde b_j}\rangle=\sum_i c_{ji}\mid b_j\rangle. 
\eea

In the next step, $V^c$ is transformed from $\mid a\rangle$ basis to $\mid b\rangle$ basis, which leads to off-diagonal terms. Finally, $V^c$ is transformed from $\mid b\rangle$ basis to $\mid{\tilde b}\rangle$ basis using $c_{ji}$ coefficients.
During the transformations, the projection
of the total angular momentum $M_F = m_{f_1} + m_{f_2}$ is conserved. The quantization axis is chosen to be along the direction of the magnetic field.

 $^6$Li has nuclear and electronic spin $i=1$ and $=1/2$, respectively. Here, we take five channels to describe  $s$-wave Feshbach resonance near 832 G following the work by \mbox{Chin et al \cite{mfr}.} These five channels are listed in Table \ref{channel}. 
In the left hand panel of Figure \ref{fig:pot+tmat}, we plot five diagonal potentials in short range regime as a function of $r$ for a particular value of the magnetic field $B=832.1$ G. In the inset of the figure, we show the asymptotic long-range part of the potentials for the chosen five channels.  
The energy of the said channels increases from \textquoteleft ab\textquoteright  to \textquoteleft de\textquoteright as a function of  $B$ \cite{mfr}. The channel \textquoteleft ab\textquoteright is open, and the other four channels are closed. 
For the open channel, we consider inward propagation from the asymptotic region $r=2000$ $a_0$, and for closed channels, we start propagation from $r=150$ $a_0$.  For this system, the matching point is chosen at $r_m \sim 22 $ $a_0$. In the right-hand panel of Figure \ref{fig:pot+tmat}, we show  $|T|^2$ as a function of energy for two values of magnetic fields 832.1 and 850 G. For the field $B=850$ G, there is hardly any effect of resonance, unlike that for $B=832.1$ G, at which a Feshbach resonance occurs as energy decreases below 1 $\mu$K. 

In passing, we verify whether we can really neglect the off-diagonal potential terms for $r>r_m$. For this, we evaluate the $\lambda$ parameter as defined in the preceding subsection for the lowest two channels, that is, the lowest open channel and the lowest closed channel. We find $\lambda \simeq 0.01$. So, we can reasonably neglect the off-diagonal terms for $r > r_m $.


\section{  \boldmath{\textbf{$d$}}-Wave Feshbach Resonance of $^{87}$Rb + $^{85}$Rb System}\label{sec:level8}

For higher partial-wave ($\ell > 0$) scattering in the presence of an external magnetic field, $m_F$ no longer remains a good quantum number, but 
only the projection $M_J$ on the quantization axis of the total angular momentum ${\vec J}$ is conserved. The asymptotic or uncoupled basis $\mid f_1 m_1, f_2 m_2; \ell m_{\ell} \rangle$ can be expressed in terms of the coupled basis in the following way 
\bea 
\hspace{-0.2in}\mid f_1 m_1, f_2 m_2; \ell m_{\ell} \rangle &=& \sum_{f} \langle f m_f \mid f_1 m_1, f_2 m_2 \rangle \mid f m_f \rangle \mid \ell m_{\ell} \rangle.\nonumber
\eea

The 
 matrix element of the central potential is given by
 
\begin{equation}
\begin{aligned}
&\langle f_1 m_1, f_2 m_2; \ell m_{\ell} \mid \hat{V}^{c}
\mid f_1' m_1', f_2' m_2'; \ell' m_{\ell}' \rangle  \\
&\quad =
\delta_{\ell \ell'} \, \delta_{m_{\ell} m_{\ell}'}
\sum_{f,f'} 
\langle f' m_f' \mid f_1' m_1', f_2' m_2' \rangle
\langle f_1 m_1, f_2 m_2 \mid f m_f \rangle  \\
&\qquad\hspace{-0.7in} \times
\sum_{S,I,M_S,M_I}
V_S\,
\langle (f_1 f_2) f m_f \mid S M_S, I M_I \rangle
\langle S M_S, I M_I \mid (f_1' f_2') f' m_f' \rangle,
\end{aligned}
\end{equation}
 where ${f}_1 + {f}_2 = {f}$, ${f}_1^{\prime} + {f}_2^{\prime} = {f}^{\prime}$, $m_1 + m_2 = m_f$
 and $m_1' + m_2' = m_f'$. Here
\bea
\langle S M_S, I M_I \mid (f_1 f_2) f m_f \rangle =  \langle S M_S, I M_I \mid f  m_f \rangle \nonumber\\ 
\sqrt{(2 f_1 + 1)(2 f_2 + 1)(2 S + 1)(2 I + 1)}\nonumber\\
\ninej{s_1}{i_1}{f_1}{s_2}{i_2}{f_2}{S}{I}{f}
\left( \frac{1+(1-\delta_{f_1 f_2})(-1)^{S+I+\ell}}{\sqrt{2-\delta_{f_1 f_2}}}\right).
\eea 

Here, 
 $\langle S M_S; I M_I \mid f  m_f \rangle$ is a Clebsch--Gordan coefficient, and the quantity in curly brackets is known as the $9j$-symbol. Here, $m_1 + m_2 = m_f = m'_1 + m'_2 = M_S + M_I$. 

From the effective-range expansion, the generalized scattering length for $\ell$-th partial wave can be expressed as $a_{\ell}=-\tan\delta_\ell/k^{2\ell+1}$ in the limit $E\rightarrow 0$. For $s$- and $p$-waves, the quantities $a_{\ell=0}$  and $a_{\ell=1}$ correspond to the scattering length and scattering volume, respectively. We define the dimensionless $d$-wave scattering length ${a_d}=a_\ell/\beta^5$, where $\beta=\left(2\mu C_6/\hbar^2\right)^{1/4}$ is  the characteristic length scale.

\begin{table}[b]
\caption{Six asymptotic channels for the $d$-wave Feshbach resonance of $^{87}$Rb and $^{85}$Rb. $f_1$ and $f_2$ represent the hyperfine quantum numbers of atoms $^{87}$Rb and $^{85}$Rb, respectively. }
\begin{ruledtabular}
\begin{tabular}{c c c } 
 \text{Channels} & {\text{$(f_1, f_2)$}} & {\text{$(m_{f_1}, m_{f_2})$}}  \\
 \hline

 1 & $(2, 3)$ & $(-2, -1)$  \\ 
 2 & $(2, 3)$ & $(-1, -2)$  \\
 3 & $(1, 3)$ & $(-1, -2)$  \\
 4 & $(2, 2)$ & $(-2, -1)$  \\ 
 5 & $(1, 3)$ & $(0, -3)$  \\ 
 6 & $(1, 2)$ & $(-1, -2)$  \\ 
\end{tabular}
\label{channel_85Rb87Rb}
\end{ruledtabular}
\end{table}

From the above consideration, we wish to calculate the $d$-wave Feshbach resonance in an ultracold mixture of $^{87}$Rb and $^{85}$Rb, studied experimentally by You's group \cite{you2017}. The $d$-wave Feshbach resonance may arise from two different mechanistic pathways. In the first pathway, the Feshbach resonance occurs mainly due to the coupling between an $\ell=0$ $(\ell=2)$ open channel and an $\ell'=2$ $(\ell'=0)$ closed channel, fulfilling the selection rule $\Delta \ell=2$.  In the second pathway, the $d$-wave Feshbach resonance occurs for 
$\Delta \ell=0$ for which there is a direct coupling of an open channel having $\ell=2$ with several other closed channels with $\ell'=2$. Experimentally, the atoms are initially prepared in the hyperfine channel $^{87}$Rb$\mid f=1, m_f=-1\rangle$ + $^{85}$Rb$\mid f=2, m_f=-2\rangle$ \cite{you2017}. For $^{85}$Rb, the nuclear spin is $i_1=5/2$ and for  $^{87}$Rb, it is $i_2=3/2$. For our numerical
computation, we consider only one open channel with $f_{85}=2$, $m_{f_{85}}=-2$, $f_{87}=1$,
$m_{f_{87}}=-1$, $\ell=2$, that is, the channel in which the atoms are initially
prepared. We consider several closed channels from the higher manifold. We have found that six-channel calculations with one open and five closed channels yield a $d$-wave Feshbach resonance close to the experimental value \cite{you2017}. These chosen channels are listed in Table \ref{channel_85Rb87Rb} with (1)--(5) being closed channels and (6) as an \mbox{open channel.}

\begin{figure}
\begin{center}
\includegraphics[width=0.5\textwidth]{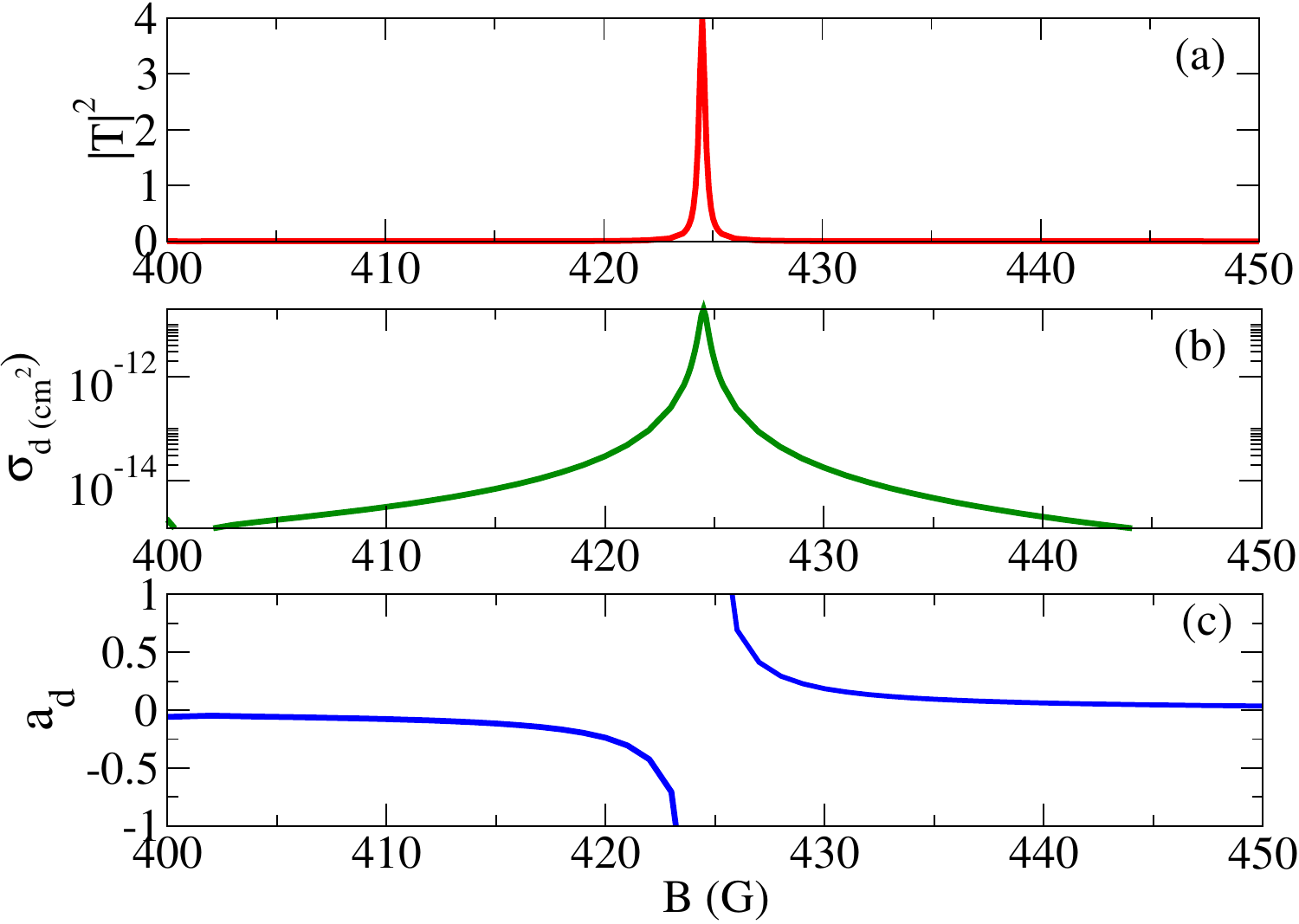}
\caption{ Shown are the $|T|$$^2$, $\sigma_d$, and the dimensionless $d$-wave scattering length $a_d$ of $^{87}$Rb + $^{85}$Rb as a function of $B$ in the panels (\text{a}--\text{c}), respectively. }
\label{T_D_ad}
\end{center}
\end{figure}

For the open channel, we start inward propagation from a large separation ($r\ge 1000 a_0$) with an asymptotic solution.
For closed channels, we choose our starting point for inward propagation nearly at $r\simeq$140 
$a_0$. We perform outward propagation from $r=9.46$a$_0$ to $r_m$ in $6\times6$ matrix form and match with numerically calculated quantum defect function using two-point matching procedure. Our matching point lies nearly at $r_m=40$ a$_0$.

\begin{figure}[t]
  \centering
  \includegraphics[width=0.5\textwidth]{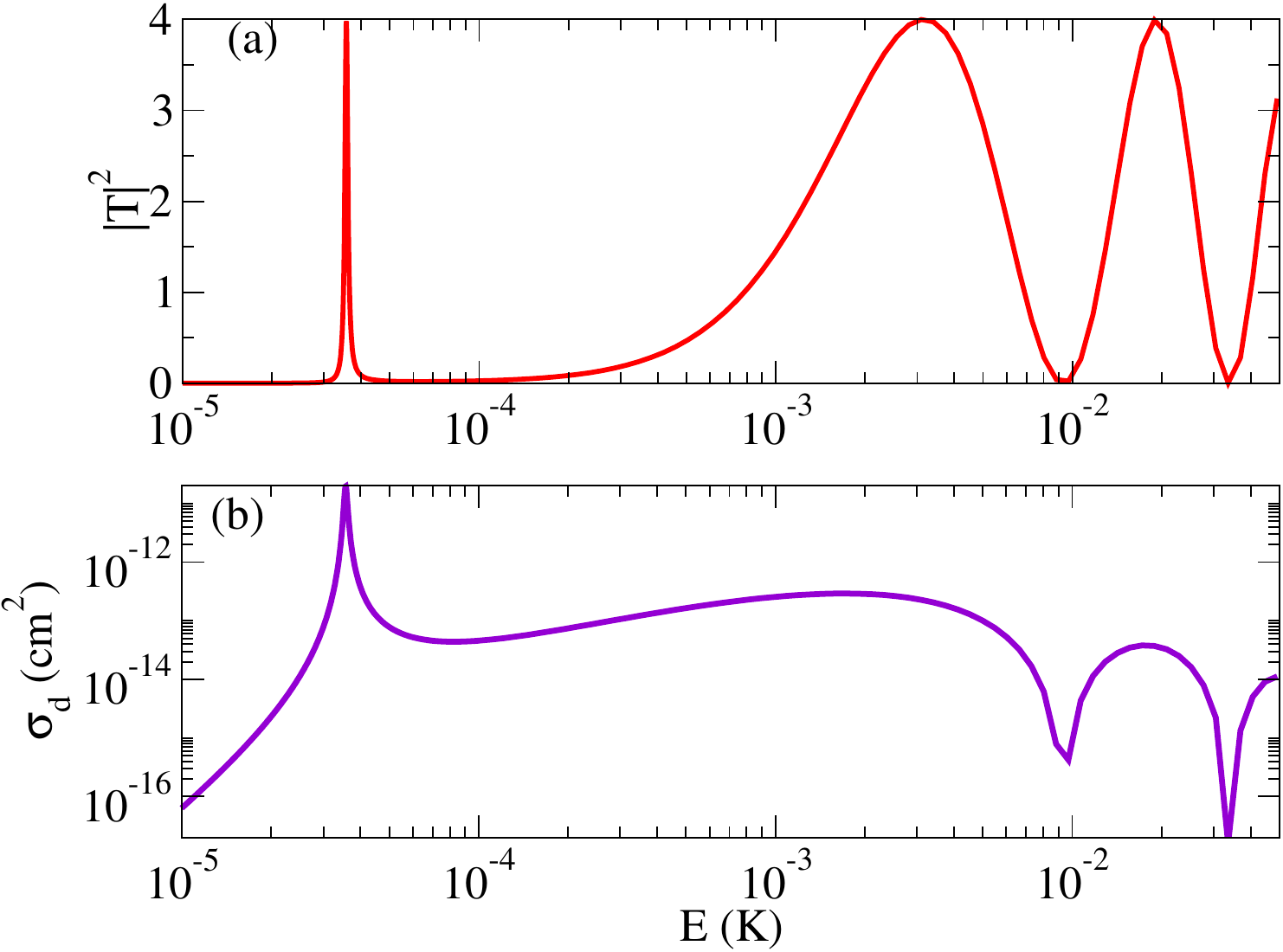}
   \caption{The variation of $|T|$$^2$ and $\sigma_d$ as a function of $E$ in the panels (\text{a}) and (\text{b}), respectively.}
  \label{sigma_SS}
\end{figure}

Next, to study the effects of the second-order spin--spin interaction on the $d$-wave resonance,  we include spin--spin interaction terms.  
The Hamiltonian for the spin--spin interaction can be expressed as  
\bea
H_{ss} = - \frac{\alpha^2}{\sqrt{6} R^3} \sum_{q=-2}^{2} (-1)^q \hat{C}^{(2)}_q \Sigma^{(2)}_{-q} 
\eea
where $\alpha$ is the fine structure constant, $\hat{C}^{(2)}_q$ is the reduced spherical harmonics, and $\Sigma^{(2)}_{-q}=(s_1\otimes s_2)^2_{-q}$ is a second-order tensor formed from the spin operators. 
The spin--spin matrix element between two basis states $\mid{\tilde b_j}\rangle$ and $\mid{\tilde b_{j'}}\rangle$ is given by 
\bea 
\langle {\tilde b_{j'}} \mid H_{ss} \mid {\tilde b_j} \rangle = \sum_{i i'} c^{*}_{j' i'} c_{j i} 
\langle b_{i'} \mid H_{ss} \mid  b_{i} \rangle
\eea 

Considering 
 $\mid b_i \rangle  \equiv  \mid f_1 m_1, f_2 m_2; \ell m_{\ell} \rangle $ and 
$\mid b_{i'}\rangle  \equiv  \mid f_1' m_1', f_2' m_2'; \ell' m_{\ell'} \rangle $, we have 
\vspace{-6pt}
\begin{equation}
    \begin{aligned}
     \langle b_{i'} \mid H_{ss} \mid  b_{i} \rangle &= \sum_{f, f'} \sum_{I, M_I}  \left [ \langle f m_f \mid f_1 m_1, f_2 m_2 \rangle \right ] \\ 
     &\times \left [ \langle S M_S, I M_I \mid (f_1 f_2) f m_f \rangle \right ]  \\
     &\times  \left [ \langle f_1' m_1', f_2' m_2' \mid  f' m_f' \rangle \right ]\\
     &\times \left [ \langle  (f_1' f_2') f' m_f' \mid S M_S', I M_I  \rangle \right ]\\
     &\times \left [ - \frac{\alpha^2}{\sqrt{6} R^3}\right ]  \sum_{q=-2}^{2} (-1)^q  \langle \ell' m_{\ell'} \mid \hat{C}^{(2)}_q \mid \ell 
m_{\ell} \rangle\\
    &  \times \langle S M_S' \mid \Sigma_{-q}^{(2)} \mid S M_S \rangle 
    \end{aligned}
\end{equation}

Here,  
 $M_S = m_f + m_{\ell} - m_I$ and $M_S' = m_f + m_{\ell'} - m_I'$. Only for the triplet state ($S=1$), the matrix element is nonzero. This follows the selection rules $\Delta \ell = 0, \pm 2$ and also fulfills the criteria: $m_{\ell'} = m_{\ell} + q$ and $M_S' = M_S - q$. 
  In our calculation, we incorporate the spin--spin interaction in matrix form up to the matching point $r_m$ and beyond $r_m$, and $H_{ss}$ is included in the diagonal channels only.

\section{  Results and Discussions} \label{sec:level9}
\begin{figure}[t]
  \centering
  \includegraphics[width=0.5\textwidth]{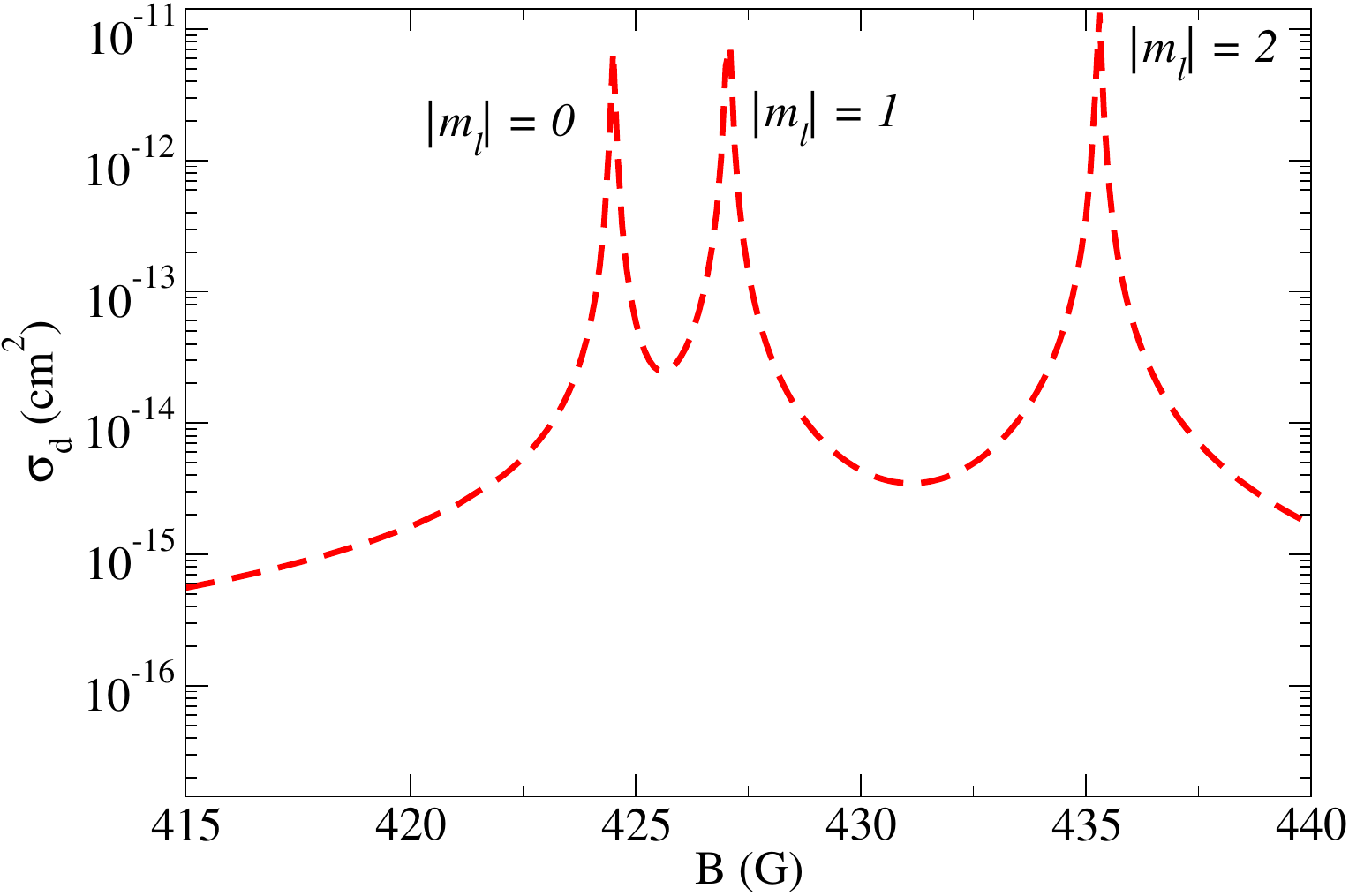}
   \caption{The variation of the $\sigma_d$ as a function of $B$ illustrates the triplet structure of the $d$-wave Feshbach resonance.}
  \label{split}
\end{figure}

We first present results of our six-channel calculations of $d$-wave Feshbach resonance without including spin--spin interactions, as shown in Figures \ref{T_D_ad} and \ref{sigma_SS}. The effects of spin--spin interactions on the Feshbach resonance are discussed subsequently in Figure \ref{split}.  The squared $T$-matrix element $\mid$T$\mid^2$, the $d$-wave scattering cross section $\sigma_d$, and the dimensionless $d$-wave scattering length $a_d$ as a function of magnetic field $B$ at a fixed collision energy $E=37$ $\mu$K, are shown in \mbox{Figure \ref{T_D_ad}}(a), (b), (c).  As shown in \mbox{Figure \ref{T_D_ad}a,} the quantity $\mid$T$\mid^2$ reaches its maximum value 4 near the magnetic field 424.1 G, indicating that the scattering phase shift passes through $\pm\pi/2$ in the vicinity of this field. This feature is a clear signature of a scattering resonance.  The corresponding resonant variation of the $d$-wave scattering cross section $\sigma_d$ is shown in \mbox{Figure \ref{T_D_ad}b,} while \mbox{Figure \ref{T_D_ad}c} exhibits the divergent behavior of the $d$-wave scattering length $a_d$ near $B = 424.1~\mathrm{G}$, further confirming the existence of a $d$-wave resonance. We also plot $|T|^2$ as a function of the collision energy in Figure \ref{sigma_SS}a at the magnetic field $B = 424.1~\mathrm{G}$. We observe that $|T|^2$ again attains its maximum value of $4$ near the energy $E = 37~\mu\mathrm{K}$. The corresponding resonant behavior of the $d$-wave scattering cross section $\sigma_d$ as a function of energy is shown in \mbox{Figure \ref{sigma_SS}b.}

In our numerical calculations, the data for the ground-state singlet and triplet potentials of Rb$_2$ are taken from  Ref. \cite{strauss2010}. The value of the van der Waals coefficient reported in this reference is $C_6=4719$ a.u.  However, since the vdW coefficient is not precisely known, this parameter may be optimized in order to obtain a good agreement between numerical results and experimental observations \cite{you2018}. We find that if we slightly change it to $C_6=4740$ a.u, we get a better agreement with the experimental results.

Experimentally, a broad single-peak $d$-wave Feshbach resonance \cite{you2017} associated with the incident channel
$^{87}$Rb$\,|f=1, m_f=-1\rangle + {}^{85}$Rb$\;|f=2, m_f=-2\rangle$
was observed at a magnetic field of approximately $423~\mathrm{G}$ at a temperature of $16~\mu\mathrm{K}$. 
Two-channel quantum defect theory calculations, which employ analytical solutions of the van der Waals potential~\cite{gao1998,gao-analytical}, predict this broad resonance at $440.9~\mathrm{G}$~\cite{you2017}. 
In contrast, our six-channel calculations yield a resonance position at $424.1~\mathrm{G}$, in much closer agreement with the experimentally observed value. Experimentally, it was further found that upon lowering the temperature to $1.2~\mu\mathrm{K}$, the single broad resonance peak splits into two distinct peaks. 
With a further reduction of the temperature to $0.4~\mu\mathrm{K}$, the resonance structure evolves into three peaks. 
In addition, as the temperature is reduced to $1.2~\mu\mathrm{K}$, both resonant peaks exhibit shifts toward higher magnetic fields.

 In Figure \ref{split}, we present $\sigma_d$ as a function of $B$, illustrating the splitting of the $d$-wave Feshbach resonance due to spin–spin interactions. The effect of spin--spin interactions is manifested at a lower energy $E = 25$ $\mu$K in the form of a triplet three-peak structure in the resonance near 424.1 G. The left peak is due to the $m_\ell=0$, while the middle and the right peaks account for the $|m_\ell|= 1$ and $|m_\ell|= 2$, respectively. Although experimentally the triplet structure in the resonance appeared at a temperature of 0.4 $\mu$K, we find such triplet structure at $E=25 \mu$K, which is lower than that at which the single-peak resonance without spin--spin couplings appears. Also, our results qualitatively agree well with the experimental results in that the spin--spin interactions lead to the shifts of the resonance point towards higher magnetic fields, and the widths of the three peaks are quite narrow ($\simeq 0.1$ G). Since the widths of the three peak structures are quite narrow, the resolution of the triplet structure will be possible if the temperature is smaller than the average width (in unit of energy) of the peak structures. In our study, we have calculated the scattering properties either as a function of $B$ at a fixed energy or as a function of energy at a fixed $B$ without considering the thermal distribution of the 
 collision energies.

\section{\label{sec:level20}Conclusions}
We have developed MQDT in a complete numerical approach using the standard Numerov-Cooley algorithm and the numerical solution of the Wronskian equation. One of the primary base functions $\phi_i(r)$ is calculated by single-channel inward propagation from the asymptotic region to the short range. The other base function is calculated by numerically solving the Wronskian equation. In our method,  we select a matching point $r_m$ near a separation where the first anti-node of the function $\phi_i(r)$ for a closed channel $i$ appears as it propagates inward, and so it lies in a classically allowed region.  After making an inward propagation, an outward propagation is carried out in matrix form. During the propagation, linear independence is maintained throughout the short as well as the long-range regime. In our method, linear independence is automatically maintained since outward multichannel propagation is carried out only within the classically allowed region. Linear independence becomes a particular issue of concern when a multichannel wave function or 
matrix wave function is propagated through a classically forbidden region. We have applied our method in three cases: (i) a standard two-channel model calculation is used for describing the $s$-wave Feshbach resonance of $^{85}$Rb atoms, (ii) a five-channel calculation is performed to describe the $s$-wave magnetic Feshbach resonance of fermionic $^6$Li system near the magnetic field $B = 832.1$ G, and (iii) a six-channel calculation is carried out to calculate the $d$-wave Feshbach resonance of $^{85}$Rb-$^{87}$Rb system near the magnetic field of 424.1 G. While the first two cases are considered only to standardize our method, new results are obtained in the third case. Our results on $d$-wave Feshbach resonance reasonably agree well with the experimental observations as reported in \cite{you2017} a few years ago.

We have justified the neglect of inter-channel couplings for $r> r_m$ by verifying whether the $\lambda$-parameter is sufficiently small. Suppose $\lambda$ is smaller than unity but not too small to neglect. In that case, the effects of the residual potential element $W^{res}_{i j}$ for $r>r_m$  can readily be taken into account perturbatively by constructing 
a real space Green function using the numerical reference functions.  Since our numerical reference functions are calculated taking into account all the long-range potential terms, the Green function so calculated will be more accurate. 
Thus, using our method, one can easily obtain the complete information of a wave function throughout the entire range for both open and closed channels. As in chemical processes like PA, the information of Franck Condon factor, which is associated with the wave function of the scattering continuum, plays an important role. Therefore, all continuum-bound spectroscopy involving atom--atom \cite{jones:rmp:2006} or atom--ion systems \cite{arpita:pra:2011,dibyendu2016,Alharzali,Farjallah,Sardar_PRA,Hela2018,Farjallah_2023}  can be explained by our MQDT technique. Our method is easy to implement, and so can be applied to all sorts of realistic long-range potentials. Complete information on the multichannel wave function or density matrix is crucially important for exploring aspects of quantum information or quantum gate operation by controlled collisions of cold \mbox{atoms \cite{jaksch,mandel}. }

\appendix
\section{Numerical Method of Solving Wronskian Equation}\label{sec:level19}
Let us consider the two linearly independent (LI) solutions $\psi_i(r)$ and $\phi_i(r)$ of the following linear second order homogeneous differential equation
\bea
\label{wr0}
y''(r) + Q(r)y(r)=0
\eea

Hence, 
\bea
\label{wr1}
\psi_i''(r)+Q(r)\psi_i(r)=0
\eea
\bea
\label{wr2}
\phi''_i(r)+Q(r)\phi_i(r)=0
\eea

Multiplying 
 Equation \eqref{wr1} by $\phi_i$, Equation \eqref{wr2} by $ \psi_i$),  and subtracting the resulting equations from each other,  we get
$W'(r) = 0$,  where $W(r)=W[\phi_i,\psi_i] = \phi'_i(r)\psi_i(r)-\psi'_i(r)\phi_i(r)$ is the Wronskian between $\phi_i(r)$ and $\psi_i(r)$, implying 
\vspace{-6pt}
\bea
\label{wr3}
 \phi'_i(r)\psi_i(r)-\psi'_i(r)\phi_i(r) = C
\eea
where $C$ is a constant.

 Let us first consider two LI functions for a closed channel $i$. Suppose the function $\phi_i(r)$ has the asymptotic boundary condition $\phi_i \sim \exp(-\kappa_i r)$. We numerically calculate this function by inward integration of single-channel Schr\"{o}dinger equation using this boundary condition. So, the other LI function $\psi_i(r)$ must satisfy the boundary condition $\psi_i(r) \sim \exp(\kappa r)$ as $r \rightarrow \infty$. Therefore,  we set $C = - 2 \kappa_i$ for a closed channel. Now, the problem at hand is to solve the Equation (\ref{wr3}) for the second LI solution $\psi(r)$. 

Let $r_a$ be a point in this large $r$ regime, and the value of the first solution ($\phi_i$, say) be known at $r_a-h$, $r_a$, and $r_a+h$, with $h$ being the step size for propagation. Then $\phi'_i(r_a)$ can be calculated as
\bea
\label{wr4}
\phi'_i(r_a)=\frac{\phi_i(r_a+h)-\phi_i(r_a-h)}{2h}
\eea
and the value of the second solution at $r_a$ is 
\bea
\label{wr5}
\psi_i(r_a)=\exp(\kappa r_a)
\eea

From 
 the Wronskian Equation (\ref{wr3}), we can write
\bea
\label{wr6}
\psi'_i(r_a)=\frac{-2\kappa+\psi_i(r_a)\phi'_i(r_a)}{\phi_i(r_a)}
\eea

The 
 value of the second derivative $\psi''_i(r_a)$ is given by the Schroedinger Equation \eqref{wr0} itself as
\bea
\label{wr7}
\psi''_i(r_a)=-Q(r_a)\psi_(r_a)
\eea

Now, 
 a Taylor series expansion of $\psi$ about $r_a$ gives
\bea
\label{wr8}
\psi_i(r_a-h)=\psi_i(r_a)-h\psi'_i(r_a)+\frac{h^2}{2}\psi''_i(r_a)
\eea

Knowing 
 the value of $\psi_i(r_a)$, $\psi'_i(r_a)$, and $\psi''_i(r_a)$ from Equation \eqref{wr5}, Equation \eqref{wr6}, and Equation \eqref{wr7}, respectively, we can get $\psi_i(r_a-h)$ from Equation \eqref{wr8}. We repeat over and over the steps of Equations \eqref{wr4}--\eqref{wr8} and calculate $\psi_i$ in the desired range.  While executing the propagation, we avoid dealing with a too small or a too large number by setting the asymptotic boundary function  $\phi_i(r) = {\mathcal N} \exp(- \kappa r)$ with a judiciously chosen normalization factor ${\mathcal N}$. The integration of Wronskian equation for finding $\psi_i(r)$ may be restricted over a limited region near the outer turning point in order to avoid the appearance of a large number.
For finding  $\psi_i(r)$ by numerical integration of the Wronskian equation, it is not necessary that one should perform inward propagation 
starting from the asymptotic separation. One can instead carry out outward propagation starting from a node point of $\phi_i(r)$ in the classically allowed region. In that case, the inward propagation for $\phi_i(r)$ should  be extended beyond  the first node point counted from the outer side. However,  the matching should be done at or near the first anti-node point. 

After calculating numerical functions for closed channels, we calculate pair functions for open channels. For an open channel $i$, we set 
$C =  k_i$ and the asymptotic boundary condition  
\bea 
\phi(r) \sim \sin(k_ir-\ell\pi/2)
\eea

We 
 calculate $\phi(r)$ numerically by inward integration of the Schr\"{o}edinger equation. We calculate the other LI 
solution $\psi_r $ 
that asymptotically behaves as 
 \bea
 \psi(r)\sim  \cos(k_ir-\ell\pi/2)
 \eea 
 by solving the Wronskian equation by the same procedure as in  the case of the closed channel, but  at 
the nodes of $\phi(r)$ we set  $\psi(r) = C/\phi^{\prime}(r)$.

\begin{acknowledgments}
One of us (Dibyendu Sardar) is thankful to CSIR Government of India for a financial support. Dibyendu Sardar is thankful to John L Bohn for discussions.
\end{acknowledgments}


\begin{thebibliography}{999}
 \bibitem{balakrishnan:jcp:2016} Balakrishnan, N. Perspective: Ultracold molecules and the dawn of cold controlled chemistry. {\it J. Chem. Phys.} {\bf 2016}, \emph{145}, 150901.

 \bibitem{coldchemistry} Bell, M.T.; Softley, T.P. Ultracold molecules and ultracold chemistry. {\it Mol. Phys.} {\bf 2008}, \emph{107}, 99.
 
 \bibitem{mfr} Chin, C.; Grimn, R.; Julienne, P.S; Tiesinga, E. Feshbach resonances in ultracold gases. {\it Rev. Mod. Phys.} {\bf 2020},  \emph{82}, 1225. 
 
 \bibitem{kohler2006} K\"{o}hler, T.; G\'{o}ral, K.; Julienne, P.S.  Production of cold molecules via magnetically tunable Feshbach resonances. {\it Rev. Mod. Phys.} {\bf 2006}, \emph{78}, 1311.
 
\bibitem{ketterle:nature:1998} Inouye, S.; Andrews, M.R.; Stenger, J.; Miesner, H.J.;  Stamper-Kurn, D.M.; Ketterle, W. Observation of Feshbach resonances in a Bose–Einstein condensate. {\it Nature.} {\bf 1998}, \emph{392}, 151.


 \bibitem{bloch:rmp:2008} Bloch, I.; Dalibard, J.; Zwerger, W. Many-body physics with ultracold gases. {\it Rev. Mod. Phys.} {\bf 2008}, \emph{80}, 885.
 
 
 \bibitem{fedichev:prl:1996}  Fedichev, P.O.;  Kagan, Y.;  Shlyapnikov, G.V.; Walraven, J.T.M. Influence of nearly resonant light on the scattering length in low-temperature atomic gases. {\it Phys. Rev. Lett.} {\bf 1996}, \emph{77}, 2913. 
 
 \bibitem{lett:prl:2002}  Fatemi, F.K.; Jones, K.M.; Lett, P.D. Observation of optically induced Feshbach resonances in collisions of cold atoms. {\it Phys.
Rev. Lett.} {\bf 2002}, \emph{85}, 4462.
 
 \bibitem{deb:prl:2009} Deb, B.; Hazra, J. Manipulating higher partial-wave atom-atom interactions by strong photoassociative coupling. {\it Phys. Rev. Lett.} {\bf 2009}, \emph{103}, 023201.
 
 \bibitem{takahashi2008}  Enomoto, K.; Kasa, K.; Kitagawa, M.; Takahashi, Y. Optical Feshbach resonance using the intercombination transition. {\it Phys. Rev. Lett.} {\bf 2008} \emph{101}, 203201. 
 
 \bibitem{jones:rmp:2006}  Jones, K.M.; Tiesinga, E.; Lett, P.D.;  Julienne, P.S. Ultracold photoassociation spectroscopy: Long-range molecules and atomic scattering. {\it Rev. Mod. Phys.} {\bf 2006}, \emph{78}, 483. 
 
 \bibitem{wiener:rmp:1999} Weiner, J.; Bagnato, V.S.;  Zilio, S.; Julienne, P.S. Experiments and theory in cold and ultracold collisions. {\it Rev. Mod. Phys.} {\bf 1999}, \mbox{\emph{71}, 1.} 
 
\bibitem{deb1} Deb, B.; Agarwal, G.S. Feshbach resonance-induced Fano interference in photoassociation. {\it J. Phys. B At. Mol. Opt. Phys.} {\bf 2009}, \emph{42}, 215203.  

\bibitem{li2} Li, Y.; Feng, G.;  Wu, J.;  Ma, Jie.; Deb, B.; Pal, A.; Xiao, L.;  Jia, S. Fano effect in an ultracold atom-molecule coupled system. {\it Phys. Rev. A.} {\bf 2019}, \emph{99}, 022702.

\bibitem{zhao:pccp:2024} Zhao, H.; Sun, Z. An improved method for reactive scatterings in ultra-cold conditions using the time-dependent approach. {\it Phys. Chem. Chem. Phys.} {\bf 2024}, \emph{26}, 22790.

\bibitem{zhao:jctc:2024} Zhao, H.; Sun, Z. Theoretical development of the interaction-asymptotic region decomposition method for tetratomic reactive scattering. {\it J. Chem. Theory Comput.} {\bf 2024}, \emph{20}, 1802.


\bibitem{burke:rmp:1962} Burke, P.G.; Smith, K. The low-energy scattering of electrons and positrons by hydrogen atoms. {\it Rev. Mod. Phys.} {\bf 1962}, \emph{34}, 458.

 \bibitem{tamura} Tamura, T. Analyses of the scattering of nuclear particles by collective nuclei in terms of the coupled-channel calculation. {\it Rev. Mod. Phys.} {\bf 1965}, \emph{37}, 679.
 
 \bibitem{hutson94} Hutson, J.M. Coupled channel methods for solving the bound-state Schrödinger equation. {\it Com. Phys. Commun.} {\bf 1994}, \emph{84}, 1. 
  
  
  \bibitem{tiecke} Tiecke, T.G.;  Goosen, M.R.;  Walraven, J.T.; Kokkelmans, S.J.J.M.F. Asymptotic Bound-state Model for Feshbach Resonances. {\it Phys. Rev. A.} {\bf 2010}, \emph{82}, 042712.

\bibitem{wille} Wille, E.;  Spiegelhalder, F.M.; Kerner, G.;  Naik, D.;  
Trenkwalder, D.;  Hendl, G.;  Schreck, F.;  Grimm, R.;  Tiecke, T.G.; 
 Walraven, J.T.M.;  et al.  Exploring an Ultracold Fermi-Fermi Mixture: Interspecies Feshbach Resonances and Scattering Properties of {$^6$Li}
and $^{40}$K. {\it Phys. Rev. Lett.} {\bf 2008}, \emph{100}, 053201.
  
  \bibitem{seaton} Seaton, M.J. Quantum defect theory I. General formulation. 
 {\it Proc. Phys. Soc.} {\bf 1966}, 88, 801; Seaton, M. J. Quantum defect theory. {\it Rep. Prog. Phys.} {\bf 1983}, \emph{46}, 167.
  
  

 
  \bibitem{greene} Greene, C.; Fano, U.; Strinati, G. General form of the quantum-defect theory. {\it Phys. Rev. A.} {\bf 1979}, \emph{19}, 1485.
  
  \bibitem{greene1} Greene, C.H.;  Rau, A.R.P.; Fano, U. General form of the quantum-defect theory. II. {\it Phys. Rev. A.} {\bf 1982}, \emph{26}, 2441.
  
  \bibitem{watanable} Watanabe, S.;  Greene, C.H. Atomic polarizability in negative-ion photodetachment  {\it Phys. Rev. A.} {\bf 1980}, \emph{22}, 158.
  
  
  
 \bibitem{book} Watanabe, S.  Doubly excited states of the helium negative ion. {\it Phys. Rev. A.} {\bf 1982}, \emph{25}, 2074. 
 
\bibitem{raoult} Raoult, M.;  Balint-Kurti, G.G. Application of generalized quantum-defect theory to photodissociation processes: Predissociation of Ar-H$_2$. {\it Phys. Rev. Lett.} {\bf 1988}, \emph{61}, 2538.

\bibitem{raoult1} Raoult, M.; Balint-Kurti, G.G. Frame transformation theory for heavy particle scattering: Application to the rotational predissociation of Ar-H$_2$. {\it J. Chem. Phys.} {\bf 1990}, \emph{93}, 6508.

\bibitem{osterwalder:jcp:2004} Osterwalder, A.; Wuest, A.; Markt, F.; Jungen, C. High-resolution millimeter wave spectroscopy and multichannel quantum defect theory of the hyperfine structure in high Rydberg states of molecular hydrogen H$_2$. {\it J. Chem. Phys.} {\bf 2004}, \emph{121}, 11810. 

\bibitem{croft} Croft, J.F.E.;  Wallis, A.O.G.;  Hutson, J.M.; Julienne, P.S. Multichannel quantum defect theory for cold molecular collisions. {\it Phys.
Rev. A.} {\bf 2011}, \emph{84}, 042703.
  
\bibitem{jisha-pra} Hazra, J.;  Ruzic, B.P.; Bohn, J.L.;  Balakrishnan, N. Quantum defect theory for cold chemistry with product-quantum-state resolution.  {\it Phys. Rev. A.} {\bf 2014}, \emph{90}, 062703.

\bibitem{jisha2014} Hazra, J.;  Ruzic, B.P.; Balakrishnan, N.; Bohn, J.L. Multichannel quantum defect theory for ro-vibrational transitions in ultracold molecule-molecule collisions. {\it Phys. Rev. A.} {\bf 2014}, \emph{90}, 032711.  
  
  
\bibitem{mies} Mies, F.H. A multichannel quantum defect analysis of diatomic predissociation and inelastic atomic scattering. {\it J. Chem. Phys.} {\bf 1984}, \emph{80}, 2514.

\bibitem{mies1}  Mies, F.H.; Julienne, P.S. A multichannel quantum defect analysis of two‐state couplings in diatomic molecules.  {\it J. Chem. Phys.} {\bf 1984}, \emph{80}, 2526.
 
\bibitem{julienne} Julienne, P.S.;  Mies, F.H. Collisions of ultracold trapped atoms. {\it J. Opt. Soc. Am. B.} {\bf 1989}, \emph{6}, 2257.

 \bibitem{burke} Burke, J.P.; Greene, C.H.;  Bohn, J.L. Multichannel cold collisions: Simple dependences on energy and magnetic field. {\it Phys. Rev. Lett.} {\bf 1998}, \emph{81},
 3355.
     
\bibitem{mies2} Mies, F.H.;  Raoult, M. Analysis of threshold effects in ultracold atomic collisions. {\it Phys. Rev. A.} {\bf 2000}, \emph{62}, 012708.
      
\bibitem{mies3} Raoult, M.; Mies, F.H. Feshbach resonance in atomic binary collisions in the Wigner threshold law regime. {\it Phys. Rev. A.} {\bf 2004}, \emph{70}, 012710.

\bibitem{tiesinga} Gao, B.;  Tiesinga, E.; Williams, C.J.; Julienne, P.S. Multichannel quantum-defect theory for slow atomic collisions. {\it Phys.
Rev. A.} {\bf 2005}, \emph{72}, 042719.

\bibitem{pires}  Pires, R.;  Repp, M.;  Ulmanis, J.;  Kuhnle, E.D.;  Weidem{\"u}ller, M.; Tiecke, T.G.; Greene, C.H.; Ruzic, B.P.;  Bohn, J.L.; Tiemann, E. Analyzing Feshbach resonances: A $^6$Li-$^{133}$Cs case study. {\it Phys. Rev. A.} {\bf 2014}, \emph{90}, 012710.

\bibitem{hanna} Hanna, T.M.;  Tiesinga, E.; Julienne, P.S. Prediction of Feshbach resonances from three input parameters. {\it Phys. Rev. A.} {\bf 2009}, \emph{79}, 040701.

\bibitem{idziaszek1} Idziaszek, Z; Calarco, T.; Julienne, P.S., Simoni, A. Quantum theory of ultracold atom-ion collisions. {\it Phys. Rev. A.} {\bf 2009}, \emph{79}, 010702.

\bibitem{idziaszek2} Idziaszek, Z.; Simoni, A.; Calarco, T.; Julienne, P.S. Multichannel quantum-defect theory for ultracold atom–ion collisions. {\it New J. Phys.} {\bf 2011}, \emph{13}, 083005.

\bibitem{gao-analytical} Gao, B. Solutions of the Schr{\"o}dinger equation for an attractive $1/r^6$
potential. {\it Phys. Rev. A.} {\bf 1998}, \emph{58}, 1728.


\bibitem{gao1998}  Gao, B.  Quantum-defect theory of atomic collisions and molecular vibration spectra. {\it Phys. Rev. A.} {\bf 1998}, \emph{58}, 4222.

\bibitem{gao2010} Gao, B. Universal properties in ultracold ion-atom interactions. {\it Phys. Rev. Lett.} {\bf 2010}, \emph{104}, 213201.

\bibitem{gao2009} Gao, B. Analytic description of atomic interaction at ultracold temperatures: The case of a single channel. {\it Phys. Rev. A.} {\bf 2009}, \emph{80}, 012702.

\bibitem{gao2006}  Julienne, P.S.; Gao, B.  Simple theoretical models for resonant cold atom interactions. {\it AIP Conf. Proc.} {\bf 2006}, \emph{869}, 261.

\bibitem{ruzic} Ruzic, B.P.;   Greene, C.H.; Bohn, J.L.  Quantum defect theory for high-partial-wave cold collisions. {\it Phys. Rev. A.} {\bf 2013}, \emph{87}, 032706.

\bibitem{you2017} Cui, Y.;  Shen, C.; Deng, M.; Dong, S.; Chen, C.; Lu, R.; Gao, B.; Tey, M.K.; You, L. Observation of Broad $d$-Wave Feshbach Resonances with a Triplet Structure {\it Phys. Rev. Lett.} {\bf 2017}, \emph{119}, 203402.

\bibitem{li} Li, M.; Gao, B. Quantum-defect theory of resonant charge exchange. {\it Phys. Rev. A.} {\bf 2012}, \emph{86}, 012707.


\bibitem{gao} Gao, B. Quantum-defect theory for $-1/r^4$-type interactions. {\it Phys. Rev. A.} {\bf 2013}, \emph{88}, 022701.


\bibitem{johnson1977} Johnson, R.B. New numerical methods applied to solving the one-dimensional eigenvalue problem. {\it J. Chem. Phys.} {\bf 1977}, \emph{67}, 4086.



\bibitem{debbook2026}  Deb, B. Low Energy Collision Physics: Applications to Cold Atoms; IOP Publishing: Bristol, UK
, 2026.


\bibitem{nygard}  Nygaard, N.;  Schneider, B.I.; Julienne, P.S. Two-channel
-matrix analysis of magnetic-field-induced Feshbach resonances. {\it Phys. Rev. A.} {\bf 2006}, \emph{73}, 042705.
  

\bibitem{burnett} K\"{o}hler, T.; Gasenzer, T.; Julienne, P.S.; Burnett, K. Long-range nature of Feshbach molecules in Bose-Einstein condensates. {\it Phys. Rev. Lett.} {\bf 2003}, \emph{91}, 230401.

\bibitem{strauss2010} Strauss, C.; Takekoshi, T.; Lang, L.; Winkler, K.; Grimm, R.; Denschlag, J.H.; Tiemann, E. Hyperfine, rotational, and vibrational structure of the $a~^3\Sigma^+_u$ state of $^{87}$Rb$_2$. {\it Phys. Rev. A.} {\bf 2010}, \emph{82}, 052514.

\bibitem{arfken}  Weber, H.J.;  Arfken, G.B. \textit {Essential Mathematical Methods for Physicists}; Academic Press: Cambrdge, MA, USA, 2003.


\bibitem{burke1999theoretical} Burke, J.P., Jr. Theoretical Investigation of Cold Alkali Atom Collisions. Ph.D. Thesis, University of Colorado Boulder, CO, \mbox{USA, 1999. }

 
\bibitem{you2018} Cui, Y.; Deng, M.; You, L.; Gao, B.; Tey, M.K. Broad Feshbach resonances in ultracold alkali-metal systems. {\it Phys. Rev. A.} {\bf 2018}, \emph{98}, 042708.
 
 
\bibitem{arpita:pra:2011} Rakshit, A.; Deb, B. Formation of cold molecular ions by radiative processes in cold ion-atom collisions. {\it Phys. Rev. A.} {\bf 2011}, \emph{83}, 022703.

\bibitem{dibyendu2016}  Sardar, D.; Naskar, S.; Pal, A.; Berriche, H.; Deb, B. Formation of a molecular ion by photoassociative Raman processes. {\it J. Phys. B.} {\bf 2016}, \emph{49}, 245202. 
 
\bibitem{Alharzali} Alharzali, N.; Sardar, D.; Mlika, R.; Deb, B.; Berriche, H. Spectroscopic properties and cold elastic collisions of alkaline-earth Mg + Mg$^+$  system. {\it J. Phys. B At. Mol. Opt. Phys.} {\bf 2018}, \emph{51}, 195201.

\bibitem{Farjallah} Farjallah, M.; Sardar, D.; El-Kork, N.; Deb, B.; Berriche, H. Electronic structure and photoassociation scheme of ultracold (MgK$^+$) molecular ions. {\it J. Phys. B At. Mol. Opt. Phys.} {\bf 2018}, \emph{52,} 045201.

\bibitem{Sardar_PRA} Sardar, D.; Naskar, S. Cold collisions between alkali metals and alkaline-earth metals in the heteronuclear atom-ion system Li + Ba$^+$. {\it Phys. Rev. A.} {\bf 2023}, \emph{107}, 043323.

\bibitem{Hela2018} Ladjimi, H.; Sardar, D.; Farjallah, M.; Alharzali, N.; Naskar, S.; Mlika, R.; Berriche, H.; Deb, B. Spectroscopic properties of the molecular ions BeX$^+$ (X=Na, K, Rb): Forming cold molecular ions from an ion–atom mixture by stimulated Raman adiabatic process. {\it Mol. Phys.} {\bf 2018}, \emph{116}, 1812.  

\bibitem{Farjallah_2023} Farjallah, M.; Sardar, D.; Deb, B.; Berriche, H. Electronic Structure, spectroscopy, cold ion–atom elastic collision properties, and Photoassociation formation prediction of the (MgCs)$^+$ molecular ion. {\it Atoms} {\bf 2023}, \emph{11}, 121.  
 

\bibitem{jaksch} Jaksch, D. Optical lattices, ultracold atoms and quantum information processing.  {\it Contemp. Phys.} {\bf 2004}, \emph{45}, 367.

\bibitem{mandel}  Mandel, O.; Greiner, M.; Widera, A.; Rom, T.; H\"{a}nsch, T.W.; Bloch, I. Controlled collisions for multi-particle entanglement of optically trapped atoms. {\it Nature} {\bf 2003}, \emph{425}, 937.

\end{thebibliography}
\end{document}